\documentclass[onecolumn,amsmath,amssymb,12pt,superscriptaddress,nofootinbib]{revtex4}
\pdfoutput=1

\usepackage[latin1]{inputenc}
\usepackage[english]{babel}
\usepackage{amssymb}
\usepackage{amsmath}
\usepackage{amsthm}
\usepackage[]{graphicx}
\usepackage[]{subfigure}
\usepackage{tensor}
\usepackage{color}
\usepackage{cancel}
\usepackage{setspace}
\usepackage{fancyhdr}
\usepackage[bookmarks,linktocpage, colorlinks=true, plainpages = false, citecolor = blue,  linkcolor=blue, urlcolor = blue, filecolor = blue]{hyperref}

\usepackage{physics}
\usepackage{tikz}

\definecolor{bulkgray}{RGB}{135,137,140}
\definecolor{topgray}{RGB}{210,211,213}

\tikzset{
  bulk/.style={
    fill=bulkgray,
    draw=black,
    line width=1.6pt
  },
  top/.style={
    fill=topgray,
    draw=black,
    line width=1.2pt
  },
  hidden/.style={
    draw=black,
    line width=1.4pt,
    dashed
  }
}

\newcommand{\nobdy}{\tikz[baseline=-0.5ex, scale=0.25]{
\path[fill=gray,draw=black,thick]
  (-2.5,2)
    .. controls (-1.5,1.2) .. (-1,-1) 
    .. controls (-0.5,-2.3) and (0.5,-2.3) .. (1, -1)
    .. controls (1.5,1.2) .. (2.5,2)
    -- cycle;
    \draw[black,line width=0.8pt]
  (-1,-1)
    .. controls (-0.5,-1.2) and (0.5,-1.2) .. (1,-1);
    \draw[black,dashed,line width=0.6pt]
  (-1,-1)
    .. controls (-0.9,-0.8) and (0.9,-0.8) .. (1,-1);
    \path[top]
  (0,2) ellipse (2.5 and 0.3);}
  }

\begin{document}

\title{Connected Geometries in\\ Gravitational Path Integrals, and\\ Normalized No-Boundary Probabilities}

\author{Jean-Luc Lehners}
	\email[]{jlehners@aei.mpg.de}
	\affiliation{Max Planck Institute for Gravitational Physics \\ (Albert Einstein Institute), 14476 Potsdam, Germany}
	
\begin{abstract}
We argue that some of the puzzling features associated with the normalization of gravitational wave functions can be substantially alleviated by restricting the path integral to a sum over connected, allowable geometries. This prescription gives sensible results when applied to classical transitions and it provides non-trivial no-boundary probabilities, in line with old expectations. We highlight the additional impact of both boundary conditions and integration contours on these issues.
\end{abstract}

\maketitle


\tableofcontents

\section{The problem} \label{sec:problem}

It is fair to say that the wave function of the universe remains an insufficiently understood concept. Yet it is also clear that the motivation to study it is high, as it lies directly at the intersection of quantum mechanics and general relativity. Wave functions involving gravity and spacetime arise in a straightforward manner when quantizing gravity semi-classically, either via canonical methods \cite{Misner:1973prb} or via a generalization of Feynman's path integral approach~\cite{Hawking:1980gf}. We will mainly be interested in the latter setting here, where quantum gravitational amplitudes can be calculated by path integrals that include gravity. Typically, such amplitudes are not normalized and their interpretation often remains rather heuristic. It is clear however that one would like to do better -- for instance because some of the most interesting theories for the beginning of our universe, in particular the Hartle-Hawking no-boundary proposal \cite{Hartle:1983ai,Lehners:2023yrj}, are phrased precisely in this language.

In a recent paper \cite{Abdalla:2026mxn}, the standard quantum mechanical normalization (reviewed in appendix \ref{sec:norm}) was applied to transitions from a no-boundary initial state to a large late-time universe, and it was argued that all such normalized amplitudes are essentially equal to unity, up to non-perturbatively suppressed factors. If true, this would imply that all states are parallel to the Hartle-Hawking no-boundary state and that, consequently, the no-boundary proposal makes no meaningful predictions. 

We will first review the basic argument here, then show that the argument cannot be right, as it leads to problematic results in much more mundane situations. Then we will explore alternative prescriptions for the normalization, with the aim of finding a more constructive definition. As we will see, the most promising approach involves the restriction to geometries that are allowable (in the sense of satisfying the Kontsevich-Segal-Witten (KSW) criterion \cite{Kontsevich:2021dmb,Witten:2021nzp}) and thus connected in the bulk (though they might have multiple disconnected boundaries), see section \ref{sec:connect}. Another important aspect is that the gravitational path integral is highly sensitive to both the choice of boundary conditions and integration contours. We will illustrate this with simple examples in section \ref{sec:mini}.

According to the Born rule (extended to gravity), the probability to reach a late-time large-universe boundary configuration $|B \rangle,$ starting from the no-boundary state $|\varnothing\rangle,$ is given by the modulus squared of the normalized transition amplitude,
\begin{align}
    P_{|\varnothing \rangle \to |B \rangle} \equiv \frac{|\langle B| \varnothing \rangle|^2}{\langle B|B \rangle \langle \varnothing | \varnothing \rangle} \,,\label{nobdyprob}
\end{align}
where transition amplitudes are calculated using the gravitational path integral, e.g.
\begin{align}
    \langle B| \varnothing \rangle = \int_\varnothing^B [Dg][D\phi] \, e^{iS[g,\phi]/\hbar}\,, 
\end{align}
where one integrates a phase (given by the action divided by the quantum of action) over all metrics $g$ and matter configurations $\phi$ that interpolate between the initial and final boundary conditions. (As we will stress later, the value of the integral depends quite sensitively on the chosen integration contours.)

The claim in \cite{Abdalla:2026mxn} is that the expression \eqref{nobdyprob} is approximately equal to one, up to exponentially small factors. The argument is most easily understood pictorially, in the style of~\cite{Abdalla:2026mxn}. We will work at the semi-classical level, i.e. in the limit of small $\hbar.$ Then the path integral is well approximated by its saddle point contributions (now setting $\hbar=1$),
\begin{align}
\langle B_2 | B_1 \rangle = \int_{B_1}^{B_2} [Dg][D\phi] \, e^{iS[g,\phi]} \approx \sum_\sigma e^{iS_\sigma}\,,   
\end{align}
where $S_\sigma$ is calculated from (in general complex) solutions to the classical equations of motion satisfying the boundary conditions $B_1$ and $B_2.$ Here $\sigma$ labels these solutions. For our purposes, we can ignore a sub-leading prefactor that arises from integration over fluctuations around the saddle point. In \cite{Abdalla:2026mxn}, contributions from all saddle points are included, without restriction, and we will take the same approach in this section.

Let us start with the vacuum amplitude, which is also sometimes called the ``vacuum persistence''. It will receive contributions from geometries that have no boundary, but that can have arbitrary topology and can have multiple disconnected components. If we denote the topology by a discrete label $a,$ then the contribution arising from connected manifolds is given by 
\begin{align}
    \sum_a e^{iS_a} \equiv \sum_a Z_a\,.
\end{align}
The simplest such contribution is that of a sphere -- in a theory of gravity coupled to a positive cosmological constant $\Lambda,$ this contribution gives the term $e^{24\pi^2/\Lambda}$ (to leading order), see also appendix \ref{sec:dS}. Terms with a more complicated topology tend to be suppressed. There may however also be disconnected components, consisting of separate connected parts that are not linked in spacetime to each other (such as two independent spheres, for example). The contribution of $n$ such disconnected components is then given by $\frac{1}{n!}(\sum Z_a)^n,$ where the factorial factor compensates for the number of permutations of disconnected components. Allowing for an arbitrary number of disconnected components then exponentiates the sum ($\sum_{n=0}^\infty x^n/n! = e^x$), with the schematic end result
\begin{align}
  \langle \varnothing|\varnothing \rangle &= \exp \left(
  \tikz[baseline=-0.5ex, scale=0.25]{
    \path[fill=gray,draw=black,thick]
      (0,0) ellipse (2 and 1.75);
      \draw[black,line width=0.8pt]
  (-2,0)
    .. controls (-0.9,-0.5) and (0.9,-0.5) .. (2,0);
    \draw[black,dashed,line width=0.6pt]
  (-2,0)
    .. controls (-0.9,0.5) and (0.9,0.5) .. (2,0);
  } \, + \, 
  \tikz[baseline=-0.5ex, scale=0.25]{
    \path[fill=gray,draw=black,thick]
      (0,0) ellipse (2 and 1.75);
    \path[fill=white]
    (0,0) ellipse (0.5 and 0.2);
    \draw[black,line width=0.8pt]
  (-0.7,0.1)
    .. controls (-0.4,-0.3) and (0.4,-0.3) .. (0.7,0.1);
    \draw[black,line width=0.6pt]
  (-0.5,-0.02)
    .. controls (-0.3,0.3) and (0.3,0.3) .. (0.5,-0.02);
  } \, + \, \cdots \, + \, 
  \tikz[baseline=-0.5ex, scale=0.25]{
    \path[fill=gray,draw=black,thick]
      (0,0) ellipse (2 and 1.75);
      \draw[black,line width=0.8pt]
  (-2,0)
    .. controls (-0.9,-0.5) and (0.9,-0.5) .. (2,0);
    \draw[black,dashed,line width=0.6pt]
  (-2,0)
    .. controls (-0.9,0.5) and (0.9,0.5) .. (2,0);
    \path[fill=gray,draw=black,thick]
      (4.2,0) ellipse (2 and 1.75);
      \draw[black,line width=0.8pt]
  (-2+4.2,0)
    .. controls (-0.9+4.2,-0.5) and (0.9+4.2,-0.5) .. (2+4.2,0);
    \draw[black,dashed,line width=0.6pt]
  (-2+4.2,0)
    .. controls (-0.9+4.2,0.5) and (0.9+4.2,0.5) .. (2+4.2,0);
  } \, + \cdots \, \right)\,.
\end{align}

Now we can also look at the transition from nothing to a final, large spatial boundary~$B.$ A possible geometry describing this transition is the no-boundary geometry. But it could also have a more complicated topology, and in fact could contain holes and handles of various dimensions. Moreover, each of these connected geometries can be augmented by disconnected components. It is straightforward to see that each connected component can contain the full vacuum persistence as an additional factor, so that we obtain
\begin{align}
  \langle B|\varnothing \rangle &= \exp \left(
  \tikz[baseline=-0.5ex, scale=0.25]{
    \path[fill=gray,draw=black,thick]
      (0,0) ellipse (2 and 1.75);
      \draw[black,line width=0.8pt]
  (-2,0)
    .. controls (-0.9,-0.5) and (0.9,-0.5) .. (2,0);
    \draw[black,dashed,line width=0.6pt]
  (-2,0)
    .. controls (-0.9,0.5) and (0.9,0.5) .. (2,0);
  } \, + \, 
  \tikz[baseline=-0.5ex, scale=0.25]{
    \path[fill=gray,draw=black,thick]
      (0,0) ellipse (2 and 1.75);
    \path[fill=white]
    (0,0) ellipse (0.5 and 0.2);
    \draw[black,line width=0.8pt]
  (-0.7,0.1)
    .. controls (-0.4,-0.3) and (0.4,-0.3) .. (0.7,0.1);
    \draw[black,line width=0.6pt]
  (-0.5,-0.02)
    .. controls (-0.3,0.3) and (0.3,0.3) .. (0.5,-0.02);
  } \, + \, \cdots \, + \, 
  \tikz[baseline=-0.5ex, scale=0.25]{
    \path[fill=gray,draw=black,thick]
      (0,0) ellipse (2 and 1.75);
      \draw[black,line width=0.8pt]
  (-2,0)
    .. controls (-0.9,-0.5) and (0.9,-0.5) .. (2,0);
    \draw[black,dashed,line width=0.6pt]
  (-2,0)
    .. controls (-0.9,0.5) and (0.9,0.5) .. (2,0);
    \path[fill=gray,draw=black,thick]
      (4.2,0) ellipse (2 and 1.75);
      \draw[black,line width=0.8pt]
  (-2+4.2,0)
    .. controls (-0.9+4.2,-0.5) and (0.9+4.2,-0.5) .. (2+4.2,0);
    \draw[black,dashed,line width=0.6pt]
  (-2+4.2,0)
    .. controls (-0.9+4.2,0.5) and (0.9+4.2,0.5) .. (2+4.2,0);
  } \, + \cdots\right)  \nonumber \\ & \qquad \times   \left( \tikz[baseline=-0.5ex, scale=0.25]{
\path[fill=gray,draw=black,thick]
  (-2.5,2)
    .. controls (-1.5,1.2) .. (-1,-1) 
    .. controls (-0.5,-2.3) and (0.5,-2.3) .. (1, -1)
    .. controls (1.5,1.2) .. (2.5,2)
    -- cycle;
    \draw[black,line width=0.8pt]
  (-1,-1)
    .. controls (-0.5,-1.2) and (0.5,-1.2) .. (1,-1);
    \draw[black,dashed,line width=0.6pt]
  (-1,-1)
    .. controls (-0.9,-0.8) and (0.9,-0.8) .. (1,-1);
    \path[top]
  (0,2) ellipse (2.5 and 0.3);
  } \, + \, \tikz[baseline=-0.5ex, scale=0.25]{
\path[fill=gray,draw=black,thick]
  (-2.5,2)
    .. controls (-1.5,1.2) .. (-1,-1) 
    .. controls (-0.5,-2.3) and (0.5,-2.3) .. (1, -1)
    .. controls (1.5,1.2) .. (2.5,2)
    -- cycle;
    \path[top]
  (0,2) ellipse (2.5 and 0.3);
  \path[fill=white]
    (0,-1.2) ellipse (0.5 and 0.2);
    \draw[black,line width=0.8pt]
  (-0.7,0.1-1.2)
    .. controls (-0.4,-0.3-1.2) and (0.4,-0.3-1.2) .. (0.7,0.1-1.2);
    \draw[black,line width=0.6pt]
  (-0.5,-0.02-1.2)
    .. controls (-0.3,0.3-1.2) and (0.3,0.3-1.2) .. (0.5,-0.02-1.2);
  } \, + \, \cdots \right)\,. \label{Jnothing}
\end{align}
Note that no additional exponentiation arises for the connected components. 

In a similar manner, we can now analyze the normalization factor associated with the large spatial geometry $B.$ This involves all geometries that contain $B$ as both an initial and a final surface. In a theory allowing de Sitter (dS) solutions, there are two obvious examples: the instantaneous ``evolution'' with zero elapsed time (which may have to be smeared~\cite{Abdalla:2026mxn}) and a symmetric slice of the closed dS solution, corresponding to a classical bounce. But it is also possible to have a contribution from two no-boundary geometries, constituted of one (``time-reversed'') geometry starting from $B$ and evolving to nothing, plus a geometry starting from nothing and evolving to $B.$ Furthermore, just as above, each such contribution can always contain an arbitrary number of disconnected  components, so that once again the vacuum persistence also becomes part of the amplitude,
\begin{align}
  \langle B|B \rangle &= \exp \left(
  \tikz[baseline=-0.5ex, scale=0.25]{
    \path[fill=gray,draw=black,thick]
      (0,0) ellipse (2 and 1.75);
      \draw[black,line width=0.8pt]
  (-2,0)
    .. controls (-0.9,-0.5) and (0.9,-0.5) .. (2,0);
    \draw[black,dashed,line width=0.6pt]
  (-2,0)
    .. controls (-0.9,0.5) and (0.9,0.5) .. (2,0);
  } \, + \, 
  \tikz[baseline=-0.5ex, scale=0.25]{
    \path[fill=gray,draw=black,thick]
      (0,0) ellipse (2 and 1.75);
    \path[fill=white]
    (0,0) ellipse (0.5 and 0.2);
    \draw[black,line width=0.8pt]
  (-0.7,0.1)
    .. controls (-0.4,-0.3) and (0.4,-0.3) .. (0.7,0.1);
    \draw[black,line width=0.6pt]
  (-0.5,-0.02)
    .. controls (-0.3,0.3) and (0.3,0.3) .. (0.5,-0.02);
  } \, + \, \cdots \, + \, 
  \tikz[baseline=-0.5ex, scale=0.25]{
    \path[fill=gray,draw=black,thick]
      (0,0) ellipse (2 and 1.75);
      \draw[black,line width=0.8pt]
  (-2,0)
    .. controls (-0.9,-0.5) and (0.9,-0.5) .. (2,0);
    \draw[black,dashed,line width=0.6pt]
  (-2,0)
    .. controls (-0.9,0.5) and (0.9,0.5) .. (2,0);
    \path[fill=gray,draw=black,thick]
      (4.2,0) ellipse (2 and 1.75);
      \draw[black,line width=0.8pt]
  (-2+4.2,0)
    .. controls (-0.9+4.2,-0.5) and (0.9+4.2,-0.5) .. (2+4.2,0);
    \draw[black,dashed,line width=0.6pt]
  (-2+4.2,0)
    .. controls (-0.9+4.2,0.5) and (0.9+4.2,0.5) .. (2+4.2,0);
  } \, + \cdots\right) \nonumber \\ & \qquad \times  \left( \nobdy
  \tikz[baseline=-0.5ex,rotate=180, scale=0.25,shift={(-4,0)}]{
\path[fill=gray,draw=black,thick]
  (-2.5,2)
    .. controls (-1.5,1.2) .. (-1,-1) 
    .. controls (-0.5,-2.3) and (0.5,-2.3) .. (1, -1)
    .. controls (1.5,1.2) .. (2.5,2)
    -- cycle;
    \draw[black,line width=0.8pt]
  (-1,-1)
    .. controls (-0.5,-1.2) and (0.5,-1.2) .. (1,-1);
    \draw[black,dashed,line width=0.6pt]
  (-1,-1)
    .. controls (-0.9,-0.8) and (0.9,-0.8) .. (1,-1);
    \path[top]
  (0,2) ellipse (2.5 and 0.3);}
   \, + \, \nobdy \tikz[baseline=-0.5ex, rotate=180, scale=0.25]{
\path[fill=gray,draw=black,thick]
  (-2.5,2)
    .. controls (-1.5,1.2) .. (-1,-1) 
    .. controls (-0.5,-2.3) and (0.5,-2.3) .. (1, -1)
    .. controls (1.5,1.2) .. (2.5,2)
    -- cycle;
    \path[top]
  (0,2) ellipse (2.5 and 0.3);
  \path[fill=white]
    (0,-1.2) ellipse (0.5 and 0.2);
    \draw[black,line width=0.8pt]
  (-0.7,0.1-1.2)
    .. controls (-0.4,-0.3-1.2) and (0.4,-0.3-1.2) .. (0.7,0.1-1.2);
    \draw[black,line width=0.6pt]
  (-0.5,-0.02-1.2)
    .. controls (-0.3,0.3-1.2) and (0.3,0.3-1.2) .. (0.5,-0.02-1.2);
  } \, + \, \cdots \, + \, 
  \tikz[baseline=-0.5ex, scale=0.25]{
\path[fill=gray,draw=black,thick]
  (-2.5,2)
    .. controls (-1.5,1.2) and (-1.05,0.7) .. (-1,0) 
    .. controls (-1.05,-0.7) and (-1.5,-1.2) .. (-2.5, -2)
    .. controls (-1.5,-2.4) and (1.5,-2.4) .. (2.5,-2)
    .. controls (1.5,-1.2) and (1.05,-0.7) .. (1,0)
    .. controls (1.05,0.7) and (1.5, 1.2) .. (2.5, 2)
    -- cycle;
    \path[top]
  (0,2) ellipse (2.6 and 0.3);
  \draw[black,dashed,line width=0.6pt]
  (-2.5,-2)
    .. controls (-1.5,-1.7) and (1.5,-1.7) .. (2.5,-2);
  }\, + \cdots\right)\,. \label{JJ}
\end{align}
We can now put these contributions together to estimate the probability of generating a universe with spatial slice $B$ from nothing. This requires forming the modulus squared of \eqref{Jnothing}, which is straightforward as the complex conjugate of $\langle B | \varnothing \rangle$ is simply $\langle \varnothing|B\rangle$ and involves the same geometries in reverse. As should be evident by inspection, the vacuum persistence simply factors out of the total expression \eqref{nobdyprob} \cite{Abdalla:2026mxn}, so that we may approximate 
\begin{align}
    P^{\mathrm{disconnected}}_{|\varnothing \rangle \to |B \rangle} \equiv \frac{|\langle B| \varnothing \rangle|^2}{\langle B|B \rangle \langle \varnothing | \varnothing \rangle} \approx \frac{ \nobdy
  \tikz[baseline=-0.5ex,rotate=180, scale=0.25,shift={(-4,0)}]{
\path[fill=gray,draw=black,thick]
  (-2.5,2)
    .. controls (-1.5,1.2) .. (-1,-1) 
    .. controls (-0.5,-2.3) and (0.5,-2.3) .. (1, -1)
    .. controls (1.5,1.2) .. (2.5,2)
    -- cycle;
    \draw[black,line width=0.8pt]
  (-1,-1)
    .. controls (-0.5,-1.2) and (0.5,-1.2) .. (1,-1);
    \draw[black,dashed,line width=0.6pt]
  (-1,-1)
    .. controls (-0.9,-0.8) and (0.9,-0.8) .. (1,-1);
    \path[top]
  (0,2) ellipse (2.5 and 0.3);}
   }{ \nobdy
  \tikz[baseline=-0.5ex,rotate=180, scale=0.25,shift={(-4,0)}]{
\path[fill=gray,draw=black,thick]
  (-2.5,2)
    .. controls (-1.5,1.2) .. (-1,-1) 
    .. controls (-0.5,-2.3) and (0.5,-2.3) .. (1, -1)
    .. controls (1.5,1.2) .. (2.5,2)
    -- cycle;
    \draw[black,line width=0.8pt]
  (-1,-1)
    .. controls (-0.5,-1.2) and (0.5,-1.2) .. (1,-1);
    \draw[black,dashed,line width=0.6pt]
  (-1,-1)
    .. controls (-0.9,-0.8) and (0.9,-0.8) .. (1,-1);
    \path[top]
  (0,2) ellipse (2.5 and 0.3);}
   \,  + \, 
  \tikz[baseline=-0.5ex, scale=0.25]{
\path[fill=gray,draw=black,thick]
  (-2.5,2)
    .. controls (-1.5,1.2) and (-1.05,0.7) .. (-1,0) 
    .. controls (-1.05,-0.7) and (-1.5,-1.2) .. (-2.5, -2)
    .. controls (-1.5,-2.4) and (1.5,-2.4) .. (2.5,-2)
    .. controls (1.5,-1.2) and (1.05,-0.7) .. (1,0)
    .. controls (1.05,0.7) and (1.5, 1.2) .. (2.5, 2)
    -- cycle;
    \path[top]
  (0,2) ellipse (2.6 and 0.3);
  \draw[black,dashed,line width=0.6pt]
  (-2.5,-2)
    .. controls (-1.5,-1.7) and (1.5,-1.7) .. (2.5,-2);
  }} \approx 1\,.\label{nobdyprobdisconnected}
\end{align}
The disconnected pieces have an exponentially enhanced weighting compared to the connected ones (see appendix \ref{sec:dS}), so that the result of \cite{Abdalla:2026mxn} immediately follows: no-boundary probabilities to reach {\it any} final slice $B$ are essentially equal to unity, irrespective of the vacuum energy and irrespective of the final slice, and thus the no-boundary proposal makes no predictions at all.

It is instructive to see what happens if we apply the same prescription to a classical transition, say the evolution of the universe from now into the very near future. We can phrase this as a transition from a spatial slice with a large scale factor $B_1$ up to an incrementally increased scale factor $B_2.$ An obvious contribution to the transition $\langle B_2|B_1 \rangle$ will be the (small) classical expansion from $B_1$ to $B_2.$ Another (given that we only specify the initial and final scale factors, and not the expansion rates) will be a slice of the closed dS hyperboloid, where the universe contracts to the waist of dS space first and then re-expands up to size $B_2.$ These contributions will have a purely Lorentzian action, and thus will neither be enhanced nor suppressed. But there is also a contribution in which the universe contracts from $B_1$ to nothing, plus a disconnected component in which the universe arises from nothing and expands to size $B_2,$ so that all in all we obtain
\begin{align}
    \langle B_2|B_1 \rangle =  
    \tikz[baseline=-0.5ex, scale=0.25]{
\path[fill=gray,draw=black,thick]
  (-2.5,0.5) 
    .. controls (-2.3,0.3).. (-2,0)
    .. controls (-1.5,-0.4) and (1.5,-0.4) .. (2,0) 
    .. controls (2.3,0.3) .. (2.5,0.5)
    -- cycle;
    \path[top]
  (0,0.5) ellipse (2.6 and 0.3);
  \draw[black,dashed,line width=0.6pt]
  (-1.9,-0.1)
    .. controls (-1.5,0.2) and (1.5,0.2) .. (1.9,-0.1);
  } \, + \, 
  \tikz[baseline=-0.5ex, scale=0.25]{
\path[fill=gray,draw=black,thick]
  (-2.5,2)
    .. controls (-1.5,1.2) and (-1.05,0.7) .. (-1,0) 
    .. controls (-1.05,-0.7) and (-1.5,-1.2) .. (-2., -2)
    .. controls (-1.5,-2.4) and (1.5,-2.4) .. (2.,-2)
    .. controls (1.5,-1.2) and (1.05,-0.7) .. (1,0)
    .. controls (1.05,0.7) and (1.5, 1.2) .. (2.5, 2)
    -- cycle;
    \path[top]
  (0,2) ellipse (2.6 and 0.3);
  \draw[black,dashed,line width=0.6pt]
  (-2.,-2)
    .. controls (-1.5,-1.7) and (1.5,-1.7) .. (2.,-2);
  } \, + \, \cdots \, + \,  \tikz[baseline=-0.5ex,rotate=180, scale=0.2,shift={(0,2)}]{
\path[fill=gray,draw=black,thick]
  (-2.5,2)
    .. controls (-1.5,1.2) .. (-1,-1) 
    .. controls (-0.5,-2.3) and (0.5,-2.3) .. (1, -1)
    .. controls (1.5,1.2) .. (2.5,2)
    -- cycle;
    \draw[black,line width=0.8pt]
  (-1,-1)
    .. controls (-0.5,-1.2) and (0.5,-1.2) .. (1,-1);
    \draw[black,dashed,line width=0.6pt]
  (-1,-1)
    .. controls (-0.9,-0.8) and (0.9,-0.8) .. (1,-1);
    \path[top]
  (0,2) ellipse (2.5 and 0.3);}\nobdy
   \,  + \, \cdots
  \,.\label{classicaldisconnect}
\end{align}
Moreover the disconnected transition is strongly enhanced. Thus we would conclude that instead of just undergoing classical evolution, a highly quantum process by which the universe would vanish and reappear, with billions of years of evolution, would be much more likely. This is evidently not quite in line with what we observe on a daily basis. For this reason we must be skeptical of the proposed prescription of \cite{Abdalla:2026mxn} for calculating and normalizing the wave function. Is there a better approach?

\section{Connected and allowable geometries} \label{sec:connect}

In quantum theory we know that if we try to talk about situations that are in principle unobservable, then eventually we run into trouble. Put more positively, quantum theory is about the things that we can observe and experience. But it appears to be impossible to detect, via any measurement device in our universe, whether a separate spacetime exists. No signal from such a separate universe can reach us, essentially by definition. This argument provides the motivation to
\begin{center}
{\it restrict to connected spacetimes only.}
\end{center}
We should clarify at the outset that, contrary to what was stated in \cite{Abdalla:2026mxn}, this is not the same as removing the no-boundary state, as for example the standard no-boundary instanton is connected. Also, the sphere contribution to the vacuum amplitude remains present, while with the no-boundary state removed it would be absent.

Further support for focusing on connected spacetimes comes from the condition for allowable metrics, also known as the KSW criterion. Kontsevich and Segal originally formulated this criterion in quantum field theory defined on curved spacetimes with potentially complex metrics \cite{Kontsevich:2021dmb}. They proposed that the background complex metrics $g_{\mu\nu}$ should only be allowed if they lead to convergent path integrals upon inclusion of all possible (real valued) $p$-form matter. If we write the $p$-form field strength as $F,$ then the explicit form of the criterion is
\begin{align}
    & \Big|\exp\Big(\frac{i}{\hbar}S\Big)\Big|<1  \,\,\, \textrm{implying} \nonumber \\
    & \Re\left(\sqrt{g} \, g^{\mu_1 \nu_1} \cdots g^{\mu_{p+1} \nu_{p+1}} F_{\mu_1 \cdots \mu_{p+1}} F_{\nu_1 \cdots \nu_{p+1}}\right) > 0\,, \label{KS}
\end{align}
where $p$ must be smaller than the spacetime dimension. If one diagonalizes the metric (which can be done pointwise), $g_{\mu\nu} = \delta_{\mu\nu} \lambda_\mu$ (no summation), then the criterion simplifies and becomes the inequality \cite{Kontsevich:2021dmb}
\begin{equation}
    \Sigma \equiv \sum_{\mu=1}^D \left|\mathrm{Arg}(\lambda_{\mu})\right| < \pi\,. \label{bound}
\end{equation} 
This shows that only metrics that are ``not too complex'' are allowed. Witten \cite{Witten:2021nzp} initiated the study of this criterion in theories with dynamical gravity, and many follow-up works investigated its validity and consequences in cosmology, see for instance \cite{Lehners:2021mah,Jonas:2022uqb,Hertog:2023vot,Chen:2023hra}. 

\begin{figure}[htbp]
\centering
\includegraphics[width=0.3\textwidth]{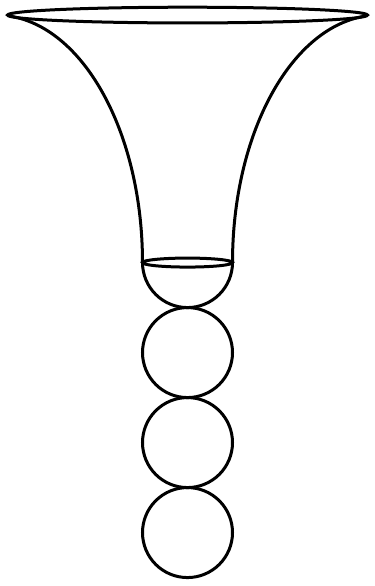}
\caption{To the no-boundary geometry one may straightforwardly attach arbitrarily many sphere contributions, which form a ``necklace'' geometry and serve as a prototype model for disconnected geometries.} \label{fig:necklace}
\end{figure}

Of particular interest to us are so-called necklace configurations, in which multiple spheres can be added to a particular solution. They serve as prototype models for disconnected contributions. A most relevant example involves no-boundary instantons. In the presence of a cosmological constant (or a scalar at a positive extremum of its potential), the $4$-dimensional Einstein equations are solved by 
\begin{equation}
    \mathrm{d}s^2 = -\mathrm{d}t^2 + a^2 \mathrm{d}\Omega_3^2\,, \qquad a(t) = \frac{1}{H} \cosh(Ht)\,, \label{metricqb}
\end{equation}
where $\mathrm{d}\Omega_3^2$ is the line element on the $3$-sphere and $\Lambda=3H^2.$ This is of course the dS solution. The standard no-boundary instanton is obtained by analytically continuing the time coordinate $t$ from $t=0$ into the complex domain, and following the path
\begin{align}
    t = -i (\tau - \frac{\pi}{2H})\quad \rightarrow \quad a(\tau)=\frac{1}{H}\sin(H\tau)\,.
\end{align}
Decreasing $\tau$ from $\pi/(2H),$ one finds a zero at $\tau=0.$ At this location, the solution caps off smoothly and the standard no-boundary solution ends there. But one may continue the $\tau$ coordinate further, reaching new maxima at $-(2n+1)\pi/(2H)$ and new zeros at $-n\pi/H,$ for integer $n.$ These correspond to additional spheres, giving a ``necklace'' configuration, see Fig.~\ref{fig:necklace}. Each additional sphere increases the weighting. One may thus wonder whether such configurations must be included.

There is a simple way of regularizing the above geometry \cite{Witten:2021nzp,Jonas:2022uqb}, which consists of going around the zeros in small semi-circles (in the complex time plane). Then it is straightforward to see that along the Lorentzian direction emanating from each $a=0$ location, and which is necessarily crossed by the regularized time contour (or by any other regularizing path, for that matter), the scale factor becomes imaginary, e.g. $\cosh(i\pi/2+H\eta) = i \sinh(H\eta)$. Thus the scale factor squared, and hence all diagonal spatial metric elements, becomes negative, so that $\Sigma = 3\pi,$ and the KSW bound \eqref{bound} is strongly violated. Thus we should discard such solutions with multiple spheres, and by extension, all disconnected spaces which, when regularized in a similar fashion, would lead to the same violation of the allowability bound.

We are now in a position to investigate how the calculation of a normalized probability changes when we restrict to connected spacetimes. Several significant changes occur. The first is that in the transition from nothing to something, there are now no disconnected vacuum factors, because they would not lead to allowable total metrics, leaving
\begin{align}
  \langle B|\varnothing \rangle &=  \left( \tikz[baseline=-0.5ex, scale=0.25]{
\path[fill=gray,draw=black,thick]
  (-2.5,2)
    .. controls (-1.5,1.2) .. (-1,-1) 
    .. controls (-0.5,-2.3) and (0.5,-2.3) .. (1, -1)
    .. controls (1.5,1.2) .. (2.5,2)
    -- cycle;
    \draw[black,line width=0.8pt]
  (-1,-1)
    .. controls (-0.5,-1.2) and (0.5,-1.2) .. (1,-1);
    \draw[black,dashed,line width=0.6pt]
  (-1,-1)
    .. controls (-0.9,-0.8) and (0.9,-0.8) .. (1,-1);
    \path[top]
  (0,2) ellipse (2.5 and 0.3);
  } \, + \, \tikz[baseline=-0.5ex, scale=0.25]{
\path[fill=gray,draw=black,thick]
  (-2.5,2)
    .. controls (-1.5,1.2) .. (-1,-1) 
    .. controls (-0.5,-2.3) and (0.5,-2.3) .. (1, -1)
    .. controls (1.5,1.2) .. (2.5,2)
    -- cycle;
    \path[top]
  (0,2) ellipse (2.5 and 0.3);
  \path[fill=white]
    (0,-1.2) ellipse (0.5 and 0.2);
    \draw[black,line width=0.8pt]
  (-0.7,0.1-1.2)
    .. controls (-0.4,-0.3-1.2) and (0.4,-0.3-1.2) .. (0.7,0.1-1.2);
    \draw[black,line width=0.6pt]
  (-0.5,-0.02-1.2)
    .. controls (-0.3,0.3-1.2) and (0.3,0.3-1.2) .. (0.5,-0.02-1.2);
  } \, + \, \cdots \right)\,. \label{Jnothing2}
\end{align}
The second change is that the vacuum persistence does not exponentiate, as there is just one connected contribution for each topology,
\begin{align}
  \langle \varnothing|\varnothing \rangle &= \left(
  \tikz[baseline=-0.5ex, scale=0.25]{
    \path[fill=gray,draw=black,thick]
      (0,0) ellipse (2 and 1.75);
      \draw[black,line width=0.8pt]
  (-2,0)
    .. controls (-0.9,-0.5) and (0.9,-0.5) .. (2,0);
    \draw[black,dashed,line width=0.6pt]
  (-2,0)
    .. controls (-0.9,0.5) and (0.9,0.5) .. (2,0);
  } \, + \, 
  \tikz[baseline=-0.5ex, scale=0.25]{
    \path[fill=gray,draw=black,thick]
      (0,0) ellipse (2 and 1.75);
    \path[fill=white]
    (0,0) ellipse (0.5 and 0.2);
    \draw[black,line width=0.8pt]
  (-0.7,0.1)
    .. controls (-0.4,-0.3) and (0.4,-0.3) .. (0.7,0.1);
    \draw[black,line width=0.6pt]
  (-0.5,-0.02)
    .. controls (-0.3,0.3) and (0.3,0.3) .. (0.5,-0.02);
  } \, + \, \cdots \,  \right)\,.
\end{align}
Moreover, the vacuum persistence does not factor out, but plays a crucial role in normalizing the probability. Finally, the normalization factor $\langle B | B \rangle$ also contrasts with the calculation in section \ref{sec:problem}, as we now only have contributions from geometries that connect the initial and final boundaries $B$, such as the instantoneous transition and a symmetric slice of the closed dS solution,
\begin{align}
  \langle B|B \rangle &=   \left(  \tikz[baseline=-0.5ex, scale=0.25]{\path[top]
  (0,0) ellipse (2.6 and 0.3);} \, + \, 
  \tikz[baseline=-0.5ex, scale=0.25]{
\path[fill=gray,draw=black,thick]
  (-2.5,2)
    .. controls (-1.5,1.2) and (-1.05,0.7) .. (-1,0) 
    .. controls (-1.05,-0.7) and (-1.5,-1.2) .. (-2.5, -2)
    .. controls (-1.5,-2.4) and (1.5,-2.4) .. (2.5,-2)
    .. controls (1.5,-1.2) and (1.05,-0.7) .. (1,0)
    .. controls (1.05,0.7) and (1.5, 1.2) .. (2.5, 2)
    -- cycle;
    \path[top]
  (0,2) ellipse (2.6 and 0.3);
  \draw[black,dashed,line width=0.6pt]
  (-2.5,-2)
    .. controls (-1.5,-1.7) and (1.5,-1.7) .. (2.5,-2);
  }\, + \cdots\right) \label{JJconnect}
\end{align}
Crucially, the enhanced contributions of geometries that vanish to nothing, only to re-emerge from nothing ``elsewhere'', are absent. 

We may then approximate the full probability as
\begin{align}
    P^{\mathrm{allowable}}_{|\varnothing \rangle \to |B \rangle} \equiv \frac{|\langle B| \varnothing \rangle|^2}{\langle B|B \rangle \langle \varnothing | \varnothing \rangle} \approx \frac{ \nobdy
  \tikz[baseline=-0.5ex,rotate=180, scale=0.25,shift={(-4,0)}]{
\path[fill=gray,draw=black,thick]
  (-2.5,2)
    .. controls (-1.5,1.2) .. (-1,-1) 
    .. controls (-0.5,-2.3) and (0.5,-2.3) .. (1, -1)
    .. controls (1.5,1.2) .. (2.5,2)
    -- cycle;
    \draw[black,line width=0.8pt]
  (-1,-1)
    .. controls (-0.5,-1.2) and (0.5,-1.2) .. (1,-1);
    \draw[black,dashed,line width=0.6pt]
  (-1,-1)
    .. controls (-0.9,-0.8) and (0.9,-0.8) .. (1,-1);
    \path[top]
  (0,2) ellipse (2.5 and 0.3);}
   }{ \left(  \tikz[baseline=-0.5ex, scale=0.25]{\path[top]
  (0,0) ellipse (2.6 and 0.3);} \, + \, 
  \tikz[baseline=-0.5ex, scale=0.25]{
\path[fill=gray,draw=black,thick]
  (-2.5,2)
    .. controls (-1.5,1.2) and (-1.05,0.7) .. (-1,0) 
    .. controls (-1.05,-0.7) and (-1.5,-1.2) .. (-2.5, -2)
    .. controls (-1.5,-2.4) and (1.5,-2.4) .. (2.5,-2)
    .. controls (1.5,-1.2) and (1.05,-0.7) .. (1,0)
    .. controls (1.05,0.7) and (1.5, 1.2) .. (2.5, 2)
    -- cycle;
    \path[top]
  (0,2) ellipse (2.6 and 0.3);
  \draw[black,dashed,line width=0.6pt]
  (-2.5,-2)
    .. controls (-1.5,-1.7) and (1.5,-1.7) .. (2.5,-2);
  }\right) \, \left( \tikz[baseline=-0.5ex, scale=0.25]{
    \path[fill=gray,draw=black,thick]
      (0,0) ellipse (2 and 1.75);
      \draw[black,line width=0.8pt]
  (-2,0)
    .. controls (-0.9,-0.5) and (0.9,-0.5) .. (2,0);
    \draw[black,dashed,line width=0.6pt]
  (-2,0)
    .. controls (-0.9,0.5) and (0.9,0.5) .. (2,0);
  }\right)_{\mathrm{vac}}} \approx \frac{ \nobdy
  \tikz[baseline=-0.5ex,rotate=180, scale=0.25,shift={(-4,0)}]{
\path[fill=gray,draw=black,thick]
  (-2.5,2)
    .. controls (-1.5,1.2) .. (-1,-1) 
    .. controls (-0.5,-2.3) and (0.5,-2.3) .. (1, -1)
    .. controls (1.5,1.2) .. (2.5,2)
    -- cycle;
    \draw[black,line width=0.8pt]
  (-1,-1)
    .. controls (-0.5,-1.2) and (0.5,-1.2) .. (1,-1);
    \draw[black,dashed,line width=0.6pt]
  (-1,-1)
    .. controls (-0.9,-0.8) and (0.9,-0.8) .. (1,-1);
    \path[top]
  (0,2) ellipse (2.5 and 0.3);}
   }{ \tikz[baseline=-0.5ex, scale=0.25]{
    \path[fill=gray,draw=black,thick]
      (0,0) ellipse (2 and 1.75);
      \draw[black,line width=0.8pt]
  (-2,0)
    .. controls (-0.9,-0.5) and (0.9,-0.5) .. (2,0);
    \draw[black,dashed,line width=0.6pt]
  (-2,0)
    .. controls (-0.9,0.5) and (0.9,0.5) .. (2,0);
  }_{\mathrm{vac}}}\,,
\end{align}
given that the classical normalization factor is of order unity.

The vacuum persistence factor now plays the role of normalizing the amplitude. Let us be more specific: the largest contribution to the vacuum amplitude will arise from a sphere solution at the lowest positive extremum of the scalar potential. Thus this overall vacuum state essentially normalizes all amplitudes. Note that it is then the case that, to leading order in $\hbar,$ the normalization factor is in fact independent of the state $B,$ and consequently the standard no-boundary probabilities, as advocated long ago by Hartle and Hawking, are recovered.

It may be of interest to look at a more explicit example. For specificity, let us assume that we have a theory with a scalar potential $V(\phi)$ that admits three extrema $V_{1,2,3}$ at positive values of the potential, such as: 
\begin{center}
\begin{tikzpicture}[domain=-2.3:2.4]

  \draw[->] (-2.3,0) -- (2.5,0) node[right] {$\phi$};
  \draw[->] (0,-0.2) -- (0,2.2) node[above] {$V(\phi)$};

  \draw[color=blue,smooth,thick]    plot (\x,{(-(\x)^4+5*(\x)^2+(\x)^3/4)/5+1/2})     node[right] {};
  \path
  (0.25,0.95) node {$V_2$}
  (-1.4,1.1) node {$V_1$}
  (1.65,1.5) node {$V_3$};
\end{tikzpicture}
\end{center}
We will also assume that the potential is such that it admits inflationary dynamics in between the potential maxima at $V_{1,3}$ and the minimum at $V_2.$ 

In line with the analysis of the KSW allowability criterion, we will take the scalar field $\phi$ to be real valued throughout. Then there are only three possible no-boundary instantons that can nucleate a universe: they all have the standard geometry of a hemisphere attached to a portion of the dS hyperboloid, with the scalar being constant at either of $\phi_{1,2,3}.$ Each of these is associated with a weighting (see appendix \ref{sec:dS})
\begin{align}
    -\mathrm{Im}[S_i] = \frac{12\pi^2}{V_i}\,, \qquad i=1,2,3\,.
\end{align}
Meanwhile, the vacuum persistence will get contributions from spheres with the scalar sitting at the same three locations, hence
\begin{align}
    \langle \varnothing | \varnothing \rangle \approx e^{24\pi^2/V_1} + e^{24\pi^2/V_2} + e^{24\pi^2/V_3} \,.
\end{align}
Of these, the contribution from $V_2$ is by far the largest, since the potential is the smallest there. If we denote the final states at large scale factor and scalar field value $\phi_{i}$ by $B_i,$ then we obtain
\begin{align}
    P^{\mathrm{allowable}}_{|\varnothing \rangle \to |B_i \rangle} \equiv \frac{|\langle B_i| \varnothing \rangle|^2}{\langle B_i|B_i \rangle \langle \varnothing | \varnothing \rangle} \approx \frac{e^{24\pi^2/V_i}}{e^{24\pi^2/V_1} + e^{24\pi^2/V_2} + e^{24\pi^2/V_3}}\,,
\end{align}
which is a sensible result. As has been discussed many times (e.g. \cite{Hertog:2013mra}), the highest likelihood for a universe arising via a no-boundary configuration resides at the lowest possible location in the potential. However, one must take into account that if such a minimum is very small, for instance residing at the currently measured value of the dark energy, then this nucleation process would give rise to an empty universe, as there would be no mechanism to populate it with stars and galaxies. Hence, adopting a weak form of the anthropic principle, what this probability distribution is really telling us is that the highest likelihood would arise for the lowest extremum that can subsequently generate the required density perturbations to produce structure in the universe. In our example this would be extremum number $1.$

Before continuing, let us note that a refinement of the present analysis will be required if one allows the scalar field to also be complex. (This might make sense in a Kaluza-Klein context, for example, because the scalar could be part of a higher-dimensional metric and thus would be allowed to be complex, though not too much \cite{Lehners:2022xds}; see also \cite{Hertog:2023vot,Lehners:2023pcn}). In that case complex no-boundary instantons would also exist, which would reach final values $\phi_f$ that are on a slope in an inflationary part of the potential \cite{Lyons:1992ua}. These would have a higher weighting than the instanton interpolating to the nearby local maximum of the potential, though they would still have a much smaller weighting than that of the lowest extremum. The problem is that the complex solutions might not contribute to the vacuum persistence, as it is unclear whether regular solutions with spherical topology and a complex, evolving scalar field exist. Thus, these solutions raise the question of whether the sum over all possible evolutions would yield a total probability of unity, or a value slightly larger. The total would not be much larger than unity, as these complex no-boundary instantons are exponentially suppressed compared to the vacuum persistence calculated above, but the question should certainly be clarified. We leave this for future work and note, for now, that the cleanest case is the one where the scalar is real valued throughout.   

We should of course also check that the restriction to connected, allowable metrics gives reasonable results when considering transitions between classical configurations. So let us again, as we did at the end of section \ref{sec:problem}, consider a transition between two large spatial slices, the second being slightly larger than the first, and which we denote by $B_{1,2}$ respectively. Then the transition amplitude between the first and second slices will include a small expanding geometry, as well as a dS bounce, plus connected contributions with a more complicated topology,
\begin{align}
    \langle B_2|B_1 \rangle =  
    \tikz[baseline=-0.5ex, scale=0.25]{
\path[fill=gray,draw=black,thick]
  (-2.5,0.5) 
    .. controls (-2.3,0.3).. (-2,0)
    .. controls (-1.5,-0.4) and (1.5,-0.4) .. (2,0) 
    .. controls (2.3,0.3) .. (2.5,0.5)
    -- cycle;
    \path[top]
  (0,0.5) ellipse (2.6 and 0.3);
  \draw[black,dashed,line width=0.6pt]
  (-1.9,-0.1)
    .. controls (-1.5,0.2) and (1.5,0.2) .. (1.9,-0.1);
  } \, + \, 
  \tikz[baseline=-0.5ex, scale=0.25]{
\path[fill=gray,draw=black,thick]
  (-2.5,2)
    .. controls (-1.5,1.2) and (-1.05,0.7) .. (-1,0) 
    .. controls (-1.05,-0.7) and (-1.5,-1.2) .. (-2., -2)
    .. controls (-1.5,-2.4) and (1.5,-2.4) .. (2.,-2)
    .. controls (1.5,-1.2) and (1.05,-0.7) .. (1,0)
    .. controls (1.05,0.7) and (1.5, 1.2) .. (2.5, 2)
    -- cycle;
    \path[top]
  (0,2) ellipse (2.6 and 0.3);
  \draw[black,dashed,line width=0.6pt]
  (-2.,-2)
    .. controls (-1.5,-1.7) and (1.5,-1.7) .. (2.,-2);
  } \, + \, \cdots 
  \,.\label{classicalconnect}
\end{align}
But this time the bothersome disconnected solution, in which the universe disappears and reappears, is excluded. Meanwhile, the normalization factors will just be two copies of \eqref{JJconnect}. Thus all leading contributions to the probability have zero weighting, and to leading semi-classical order the probability will simply be unity,
\begin{align}
    P^{\mathrm{allowable}}_{|B_1 \rangle \to |B_2 \rangle} \equiv \frac{|\langle B_2| B_1 \rangle|^2}{\langle B_1|B_1 \rangle \langle B_2 | B_2 \rangle} \approx \frac{ \left( \tikz[baseline=-0.5ex, scale=0.25]{
\path[fill=gray,draw=black,thick]
  (-2.5,0.5) 
    .. controls (-2.3,0.3).. (-2,0)
    .. controls (-1.5,-0.4) and (1.5,-0.4) .. (2,0) 
    .. controls (2.3,0.3) .. (2.5,0.5)
    -- cycle;
    \path[top]
  (0,0.5) ellipse (2.6 and 0.3);
  \draw[black,dashed,line width=0.6pt]
  (-1.9,-0.1)
    .. controls (-1.5,0.2) and (1.5,0.2) .. (1.9,-0.1);
  } \, + \, 
  \tikz[baseline=-0.5ex, scale=0.25]{
\path[fill=gray,draw=black,thick]
  (-2.5,2)
    .. controls (-1.5,1.2) and (-1.05,0.7) .. (-1,0) 
    .. controls (-1.05,-0.7) and (-1.5,-1.2) .. (-2., -2)
    .. controls (-1.5,-2.4) and (1.5,-2.4) .. (2.,-2)
    .. controls (1.5,-1.2) and (1.05,-0.7) .. (1,0)
    .. controls (1.05,0.7) and (1.5, 1.2) .. (2.5, 2)
    -- cycle;
    \path[top]
  (0,2) ellipse (2.6 and 0.3);
  \draw[black,dashed,line width=0.6pt]
  (-2.,-2)
    .. controls (-1.5,-1.7) and (1.5,-1.7) .. (2.,-2);
  } \right)^2
   }{ \left(  \tikz[baseline=-0.5ex, scale=0.2]{\path[top]
  (0,0) ellipse (2.6 and 0.3);} \, + \, 
  \tikz[baseline=-0.5ex, scale=0.2]{
\path[fill=gray,draw=black,thick]
  (-2.5,2)
    .. controls (-1.5,1.2) and (-1.05,0.7) .. (-1,0) 
    .. controls (-1.05,-0.7) and (-1.5,-1.2) .. (-2.5, -2)
    .. controls (-1.5,-2.4) and (1.5,-2.4) .. (2.5,-2)
    .. controls (1.5,-1.2) and (1.05,-0.7) .. (1,0)
    .. controls (1.05,0.7) and (1.5, 1.2) .. (2.5, 2)
    -- cycle;
    \path[top]
  (0,2) ellipse (2.6 and 0.3);
  \draw[black,dashed,line width=0.6pt]
  (-2.5,-2)
    .. controls (-1.5,-1.7) and (1.5,-1.7) .. (2.5,-2);
  }\right) \, \left(  \tikz[baseline=-0.5ex, scale=0.25]{\path[top]
  (0,0) ellipse (2.6 and 0.3);} \, + \, 
  \tikz[baseline=-0.5ex, scale=0.25]{
\path[fill=gray,draw=black,thick]
  (-2.5,2)
    .. controls (-1.5,1.2) and (-1.05,0.7) .. (-1,0) 
    .. controls (-1.05,-0.7) and (-1.5,-1.2) .. (-2.5, -2)
    .. controls (-1.5,-2.4) and (1.5,-2.4) .. (2.5,-2)
    .. controls (1.5,-1.2) and (1.05,-0.7) .. (1,0)
    .. controls (1.05,0.7) and (1.5, 1.2) .. (2.5, 2)
    -- cycle;
    \path[top]
  (0,2) ellipse (2.6 and 0.3);
  \draw[black,dashed,line width=0.6pt]
  (-2.5,-2)
    .. controls (-1.5,-1.7) and (1.5,-1.7) .. (2.5,-2);
  }\right)} \approx 1\,.
\end{align}
This is entirely reasonable, as we are discussing classical evolution. (The treatment would be slightly more involved if matter fields were also included.) Note that it is natural that we obtain both expanding and bouncing contributions -- this is a result of us fixing solely the initial and final sizes of the geometry. If the boundary conditions also contained sufficient information about the expansion rate (for instance if one were to impose a Robin condition), one or the other saddle could disappear. 

Hence we see that the restriction to connected, allowable geometries avoids the pathologies discovered in \cite{Abdalla:2026mxn}, and works well in simple settings. In the following section, we will be a little bit more precise in our definitions of gravitational path integrals, though we will still remain in the simple minisuperspace context, going beyond the pictorial approach of the present section.

\section{Dependence on boundary conditions and integration contours: minisuperspace examples} \label{sec:mini}

The question we are chiefly concerned with is: what geometries should the gravitational path integral sum over? What is clear is that the answer cannot be ``all geometries'' as this would include a great many pathological spacetimes and would render the path integral ill defined \cite{Witten:2021nzp}. As we saw in the previous section, a promising restriction is to focus on connected geometries only. But in practice this characterization is not precise enough. As we will see with simple examples, both the boundary conditions and the integration contours have a strong impact on the value of gravitational path integrals, to such an extent that one cannot really discuss such integrals without specifying both. These examples are well known (for early works and more details see \cite{Halliwell:1988ik,Halliwell:1989vu,Halliwell:1990tu}) -- our motivation for reviewing them is to re-examine them in light of the proposed restriction to connected geometries.

In practical examples (see e.g. \cite{Halliwell:1990tu}), it turns out that it is typically quite straightforward to choose the integration contours for the spatial components of the metric, such as the scale factor. By contrast, the integration contour for the lapse function is often contentious \cite{Feldbrugge:2017kzv,DiazDorronsoro:2017hti,DiazDorronsoro:2018wro,Feldbrugge:2018gin}. Historically, there have been two favored choices: either a Euclidean contour or a Lorentzian one. The Euclidean contour was initially thought to lead to better convergence properties \cite{Hartle:1983ai}, but it turns out to be plagued by the conformal factor problem, which implies that the gravitational action is unbounded from below -- in many situations of interest, this problem in fact renders the integral along an Euclidean contour divergent and ill-defined. By contrast, the Lorentzian contour leads to oscillating integrals. These are more difficult to evaluate, but enjoy better convergence properties \cite{Feldbrugge:2017mbc}. Many recent works have investigated their usage in cosmology and black hole physics, see e.g. \cite{Feldbrugge:2017kzv,Loges:2022nuw,Narain:2022msz,Matsui:2022lfj,Mondal:2023cxx,Ailiga:2023wzl,Lehners:2019ibe,Lehners:2023fud,Honda:2024aro,Honda:2024hdr,Dittrich:2024awu,Borissova:2024txs,Nomura:2025whc,Held:2026bbo,Kolanowski:2026gii,Assant:2026dca,Honda:2026wsl}. There is also the possibility of assigning a complex integration contour for the lapse, but, as we will see shortly, this can lead to physically dubious results.

Simple examples are sufficient to illustrate these arguments. We will consider transition amplitudes between spatial configurations with boundary conditions $B_{1,2},$ working with highly symmetric metrics so that we only keep as variables the lapse function $N$ and the scale factor $a,$  
\begin{align}
\langle B_2 | B_1\rangle = \int_{\cal C} \mathrm{d}N \int_{B_1}^{B_2} \mathcal{D} a \, e^{\frac{i}{\hbar} S(N,a)}  \,.
\end{align} 
The action is that of general relativity coupled to a positive cosmological constant $\Lambda$  (setting $8\pi G = 1$),
\begin{eqnarray}
S = \int_{\cal M} \mathrm{d} ^4 x  \sqrt{-g} \left( \frac{R}{2} - \Lambda \right)  + \gamma \int_{\partial \cal M} \mathrm{d}^3 y \sqrt{h}K \,, \label{LorentzianAction}
\end{eqnarray}
where $K$ denotes the extrinsic curvature of the initial/final spatial slices. The parameter $\gamma$ specifies the boundary conditions, in particular $\gamma=\pm 1$ for a Dirichlet condition (fixing $a$) or $\gamma=0$ for a Neumann condition (fixing the momentum). It is convenient to work with the scale factor squared $q\equiv a^2,$ and rescale both the lapse and the time coordinate such that the latter runs over $0 \leq t \leq 1.$ The metric then reads
\begin{eqnarray} \label{eq:metriq}
\mathrm{d}s ^2 =  - \frac{{N}^2}{q(t)} \mathrm{d}t^2 + q(t) \mathrm{d} \Omega _3 ^2 \,,
\end{eqnarray}
and this form of the metric renders the action quadratic in $q,$
\begin{equation}
\frac{S}{2\pi^2}= \int_0^1 \mathrm{d}t \left( \frac{3}{2 {N}}q\ddot{q} + \frac{3}{4 {N}}\dot{q}^2 + N(3 -  \Lambda q) \right) - \sum_{t=0,1} \gamma \frac{3}{2 {N}}q\dot{q}\,. \label{Actionq}
\end{equation} 
The classical equation of motion and constraint are
\begin{eqnarray}
\ddot{q}  &=&  \frac{2\Lambda}{3}{N}^2\,, \label{qeom} \\ \frac{3}{4 {N}^2} \dot{q}^2  &=& \Lambda q -3 \,. \label{qconstraint} 
\end{eqnarray}

Let us start with Dirichlet boundary conditions on both ends, fixing $q(t=0)\equiv q_0$ and $q(t=1)\equiv q_1.$ The path integral over the scale factor can be evaluated by shifting variables to $q=\bar{q} + Q,$ where $\bar{q}$ is a background solution satisfying \eqref{qeom} (though not \eqref{qconstraint}) and the boundary conditions, and the integral over $Q$ becoming just a fluctuation factor that we can ignore here \cite{Halliwell:1988ik}. In our case 
\begin{equation}
\label{eq:classicalsolution}
\bar{q}=\frac{\Lambda}{3}{N}^2 t^2 + \left(- \frac{\Lambda}{3}{N}^2+ q_1- q_0\right) t + q_0\,,
\end{equation}
so that the action reduces to a function of the lapse only,
\begin{equation}
\frac{S_0(N)}{2\pi^2} = {N}^3 \, \frac{\Lambda^2}{36} + {N} \left( -\frac{\Lambda}{2}(q_0+q_1) +3 \right) +\frac{1}{{N}}\left( -\frac{3}{4} (q_1-q_0)^2\right)\,. \label{eq:DDaction}
\end{equation}
Note that there is a singularity at $N=0,$ which arises because it is impossible to link two different scale factor values $q_0 \neq q_1$ in zero time. The saddle points of the path integral are thus given by
\begin{equation}
{N}_\sigma = c_1 \frac{3}{\Lambda} \left[ \left( \frac{\Lambda}{3} q_0-1\right)^{1/2} + c_2 \left( \frac{\Lambda}{3} q_1-1\right)^{1/2}\right]\,,\label{sadbuttrue}
\end{equation}
with $c_1, c_2\in\{-1,1\}$. And the value of the action at the saddle points is  
\begin{eqnarray}
\frac{S_{0,\sigma}({N}_\sigma)}{2\pi^2} =  -c_1 \frac{6}{\Lambda}  \left[ \left(\frac{\Lambda}{3} q_0 - 1\right)^{3/2} +c_2 \left(\frac{\Lambda}{3} q_1 - 1\right)^{3/2} \right]\,. \label{sadactions}
\end{eqnarray}

We may now distinguish two cases, one where we are interested in no-boundary amplitudes and one where we look at classical transitions. In the no-boundary case we set $q_0=0,$ to implement the idea that the universe starts from a vanishing geometry. For the final scale factor we may take any value $q_1 > 3/\Lambda.$ As \eqref{sadbuttrue} shows, all four saddle points are complex, due to $q_0$ being smaller than the dS waist. In fact, all four saddles have a no-boundary geometry, but there are four of them as they arise in complex conjugate and time-reversed pairs.

\begin{figure}[htbp]
\centering
\includegraphics[width=0.4\textwidth]{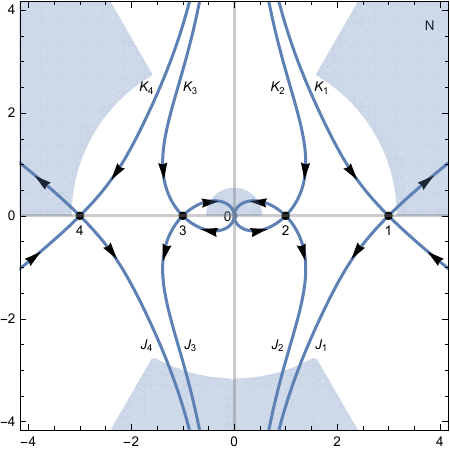}
\hspace{0.7cm}
\includegraphics[width=0.4\textwidth]{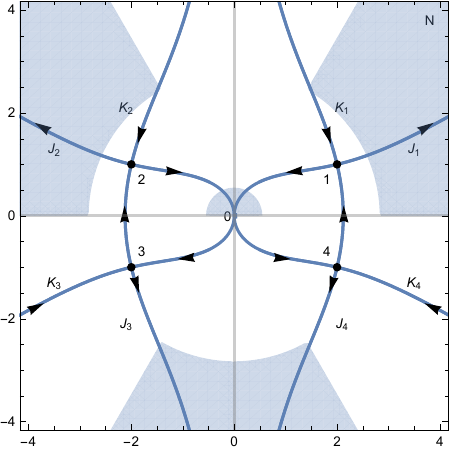}
\caption{Saddle points $N_{\pm}$ and associated flow lines in the complex plane of the lapse function, for Dirichlet-Dirichlet boundary conditions. The shaded areas indicate the asymptotic regions of convergence of the path integral. Here $\Lambda = 3, q_1 = 5.$ {\it Left:} classical starting geometry, $q_0 = 2$. {\it Right:} zero initial size, $q_0=0.$} \label{fig:DD}
\end{figure}

The flow lines emanating from the saddle points are shown in the right panel in Fig.~\ref{fig:DD}. There is an important distinction between the four saddle points: the two in the upper half $N$ plane admit perturbations with an inverse Gaussian weighting, meaning that they are highly unstable (the saddle points themselves have a negative weighting $-12\pi^2/\Lambda$). By contrast, the two saddles in the lower half plane admit stable perturbations and have the standard positive no-boundary weighting $+12\pi^2/\Lambda$. 

The shaded areas in the figure indicate the asymptotic regions of convergence of the path integral, and we can only choose integration contours that start and end in these areas. Then we can see immediately that the Euclidean contour (along the imaginary $N$ axis) does not lead to a convergent integral, and is thus mathematically meaningless. How about the Lorentzian contour? It corresponds to integrating along the real $N$ line, and passing the singularity at $N=0$ in the (convergent) upper half plane. To find out which saddle points contribute to this integral, we employ the methods reviewed in appendix \ref{sec:Picard}. The integration contour is intersected by the steepest ascent lines $K_1$ and $K_2,$ and thus only saddles $1$ and $2$ contribute. But these are the ones with unstable fluctuations, and hence we do not obtain a meaningful no-boundary amplitude.

We could resort to a complex contour, for example starting and ending at real $N$ values, but passing below the singularity at $N=0.$ This contour is intrinsically complex, due to the region of divergent weighting just below $N=0.$ This contour crosses all four steepest ascent lines, and hence it picks up both stable and unstable saddle points. It remains unclear whether the presence of the unstable saddle points is problematic or not \cite{DiazDorronsoro:2017hti,Feldbrugge:2017mbc}. We will not discuss this issue further here, but rather look at classical transitions, which provide additional context regarding the choice of integration contours. Before doing so, let us just add that the fact that the complex integration contour passes a region of divergent weighting implies that in principle geometries with arbitrarily large weighting can contribute to the integral. This includes pathological geometries such as the necklace configurations discussed earlier.

We proceed by fixing the initial size to be larger than the dS horizon, though smaller than the final configuration, $3/\Lambda < q_0 < q_1.$ Now all four saddle points are real, and the shape of the associated flow lines is shown in the left panel in Fig.~\ref{fig:DD}. Again the saddle points arise in time-reversed pairs, and hence it is enough to discuss those at positive $N,$ say. There are two saddles, one with a smaller lapse, representing a purely expanding geometry, and one with a larger lapse, representing a bouncing geometry (containing the dS waist in the middle, see appendix \ref{sec:dS}). Again, the Euclidean integral does not exist, as the regions of convergence remain unchanged. The Lorentzian contour picks up all saddle points, and is thus given by a sum of phases, with the phases specified in \eqref{sadactions}. This is the expected result, given that we just specified initial and final scale factor values, and that they can be joined via standard classical evolution.

The complex contour described above is more interesting. It can be deformed into the sum of thimbles $J_4 + J_3.$ Thus we find that the complex contour picks up only the bouncing geometries, and not those that are purely expanding between $q_0$ and $q_1.$ This is physically inconsistent. But things get worse: if we were to split the time interval into two pieces, we could have two geometries contributing: one starting from $q_0$ and then shrinking to zero size, and a second starting at zero and reaching $q_1$ -- this is of course the disconnected geometry that we have discussed repeatedly already. Since the weighting along the integration contour becomes arbitrarily positive, this geometry (as well as additional ones with more numerous disconnected pieces) would contribute to the path integral and dominate it, in even more blatant conflict with observations. Thus we should discard this type of contour. Note that the Lorentzian contour prevents such enhanced geometries from contributing. In that case, the problem was rather that one could not obtain a sensible no-boundary amplitude, but for classical transitions it works perfectly well. 

We have just seen the advantage of a Lorentzian contour. Is there a way of sticking to it, while nevertheless capturing no-boundary geometries? In fact there is, and it involves a different kind of boundary condition \cite{DiTucci:2019bui,DiTucci:2019dji,Krishnan:2016mcj}. If we momentarily go back to our starting action in \eqref{LorentzianAction}, then we recall that the boundary conditions are determined by the addition of boundary terms. The spirit of the no-boundary proposal is however that the part of the geometry that nucleates a universe should contain no boundary. This strongly suggests that one should {\it not} add any boundary term, i.e. that one should set $\gamma=0$ at $t=0.$ In that case, the variational problem changes somewhat, and in fact consistency requires that one fix the momentum rather than the initial size. We will keep a Dirichlet condition on the final slice. Thus, the quantities that we may fix are given by
\begin{align}
-\frac{3\pi^2}{N}\dot{q} \mid_{t=0} \equiv \Pi_0\,, \qquad q \mid_{t=1} \equiv q_1\,. \label{NDboundary}
\end{align}
With these boundary conditions $\bar{q}$ is given by
\begin{align}
q(t) &= H^2 N^2 (t^2-1) - \frac{N}{3\pi^2} \Pi_0 (t-1) + q_1 \,, \label{NDq}
\end{align}
and then the action for the lapse becomes (neglecting again a fluctuation factor that was evaluated in \cite{DiTucci:2020weq})
\begin{align}
S =\frac{2\pi^2}{9} \Lambda^2 N^3 -\frac{\Lambda}{3} \Pi_0 N^2 +\frac{ \Pi_0^2}{6\pi^2}N - 6\pi^2 \left(\frac{\Lambda}{3} q_1 -1\right) N  +\Pi_0 q_1 \,. \label{NDaction}
\end{align}
There are now just two saddle points, located at
\begin{align}
    N_\pm = \frac{\Pi_0}{2\pi^2 \Lambda} \pm \frac{3}{\Lambda}\sqrt{\frac{\Lambda}{3}q_1 - 1}\,,
\end{align}
and with saddle point action
\begin{align}
    S(N_\pm) = \frac{3\Pi_0}{\Lambda} + \frac{\Pi_0^3}{36\pi^4\Lambda} \mp \frac{12\pi^2\left(\frac{\Lambda}{3}q_1 - 1\right)^{3/2}}{\Lambda}\,. 
\end{align}

\begin{figure}[htbp]
\centering
\includegraphics[width=0.4\textwidth]{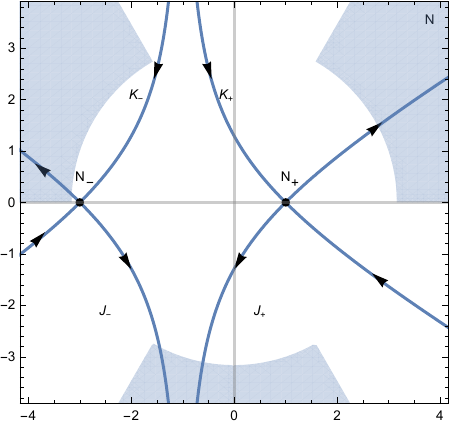}
\hspace{0.7cm}
\includegraphics[width=0.4\textwidth]{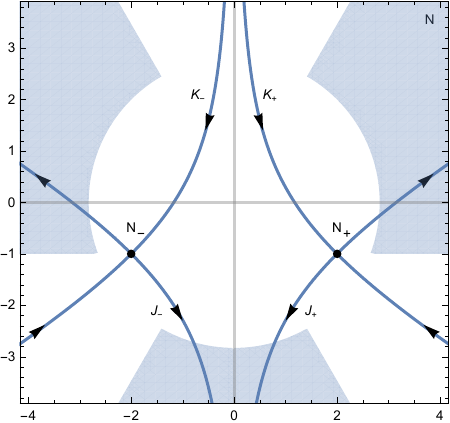}
\caption{Saddle points $N_{\pm}$ and associated flow lines in the complex plane of the lapse function, for combined Neumann-Dirichlet boundary conditions. The shaded areas indicate the asymptotic regions of convergence of the path integral. Here $\Lambda=3, q_1=5.$ {\it Left:} classical momentum condition, $\Pi_0=-6\pi^2$. {\it Right:} regularity condition on the momentum, leading to no-boundary saddle point geometries, $\Pi_0 = -6\pi^2 i$.} \label{fig:ND}
\end{figure}

We will once again look at both classical transitions and no-boundary amplitudes. This time, we start with classical transitions. These correspond to specifying a real momentum $\Pi_0 \in \mathbb{R}.$ In that case, the action and the saddle points are real, see the left panel in Fig.~\ref{fig:ND} for an example of the flow lines. The asymptotic regions of convergence at large $|N|$ remain unchanged, and thus an Euclidean lapse integral is once again impossible. The Lorentzian contour is however fine, and in fact there is no singularity on the entire real $N$ line. This contour intersects both saddle points, and can be deformed into the sum $J_- + J_+.$ The path integral will thus pick up both saddle points, and they once again correspond to a purely expanding and a bouncing geometry (the reason a bouncing geometry can be present even though we specified an initial $\dot{q}/N>0$ is that the lapse is negative there, and hence $\dot{q}$ can be negative without changing the sign of the momentum $\Pi_0$). This is once again entirely consistent. Also, since the weighting along the defining Lorentzian contour is zero, no enhanced saddle points, such as the disconnected ones that shrink to zero at some intermediate time $q(t=t_m)=0$ only to re-emerge elsewhere in order to reach the final size $q_1,$ can contribute.

Now we can also look at no-boundary transitions. What condition should we impose on the initial momentum in order to implement the no-boundary proposal? We cannot specify the initial size, rather we must specify $\Pi_0,$ as given in \eqref{NDboundary}. Guidance is provided by the constraint \eqref{qconstraint}. We would like the geometries to be regular, even when they shrink to zero size. This reduces the choice of initial momentum to
\begin{align}
    \Pi_0 = \pm 6\pi^2 i\,.
\end{align}
We need to impose the minus sign in order to obtain stable fluctuations (this is also the sign corresponding to the usual Wick rotation in quantum field theory). Fixing this sign in fact eliminates the unstable saddles, and this explains why there are only two saddles here, compared to the four we obtained with a Dirichlet condition; they are related by reversing the (arbitrary) choice of the time direction.

Now the saddle points are complex, and an example of the associated flow lines is given in the right panel of Fig.~\ref{fig:ND}. The interesting point is that the Lorentzian contour remains viable. This is because the boundary condition on the momentum is complex, and leads to a complex action, even along the contour of real $N$ values. Thus both no-boundary saddles are picked up, even though we are integrating along a Lorentzian contour.

It is worthwhile elaborating a little on this point. With the no-boundary condition $\Pi_0=-6\pi^2 i,$ the weighting of the action along the real $N$ line becomes
\begin{align}
    - \mathrm{Im}[S_0]|_{N \in \mathbb{R}} = 6\pi^2 q_1 - 2\pi^2 \Lambda N^2 \,.
\end{align}
This can be positive. In fact, the weighting is largest when $N=0,$ and given by $+6\pi^2 q_1.$ We should compare this with the expected weight of a no-boundary geometry, which is $12\pi^2/\Lambda.$ For consistency, the saddle point weight must be below that along the integration contour, so it must be true that 
\begin{align}
    q_1 > \frac{2}{\Lambda} = \frac{2}{3}\frac{1}{H^2}\,.
\end{align}
For a classical final configuration, $q_1>1/H^2,$ and thus we see that for any realistic final size the consistency condition is fulfilled.

At this level the Lorentzian contour, combined with a complex no-boundary condition, works well. Unfortunately, this is not enough to eliminate all pathological geometries by itself. The reason is that we can imagine splitting the time interval between initial and final slices into two (or many more pieces). One may then ask what the intermediate boundary condition should be, Neumann or Dirichlet? For a Dirichlet condition ($q_{intermediate}=0$), and sticking to the Lorentzian integration contour, no enhanced contribution could arise between $q_{intermediate}$ and $q_1.$ However, since we now start with a complex Neumann condition at $t=0,$ we may equally well impose such complex regularity conditions at intermediate times. Then enhanced, disconnected contributions can once again arise.

In the simple model we examined here, the equation of motion \eqref{qeom} is overdetermined when imposing two Neumann conditions -- one can however perform this calculation in a slightly enlarged minisuperspace model, for instance in the biaxial Bianchi IX model \cite{Janssen:2019sex,Feldbrugge:2018gin,Lehners:2024kus}. This calculation is of interest, because it is also the calculation that one must perform to obtain the vacuum persistence $\langle \varnothing_N | \varnothing_N \rangle$, which here must be evaluated with Neumann boundary conditions.

Hence we will generalize the metric to the following form
\begin{align} \label{metricbbix}
\mathrm{d}s^2 = - \frac{N^2}{q}\mathrm{d}t^2 + \frac{p}{4} (\sigma_1^2+\sigma_2^2) + \frac{q}{4} \sigma_3^2\,,
\end{align}
where $p(t),q(t)$ are now two separate time dependent scale factors and $\sigma_1 = \sin\psi \mathrm{d}\theta - \cos \psi \sin \theta \mathrm{d}\varphi$, $\sigma_2 = \cos \psi \mathrm{d}\theta + \sin \psi \sin \theta \mathrm{d} \varphi$, and $\sigma_3 = - (\mathrm{d}\psi + \cos\theta \mathrm{d}\varphi)$ are differential forms on the three sphere with $0 \leq \psi \leq 4 \pi$, $0 \leq \theta \leq \pi$, and $0 \leq \varphi \leq 2 \pi.$ The action \eqref{LorentzianAction}, now with $\gamma=0$ on both endpoints (since we do not want to have a boundary on either end), reduces to
\begin{align}
    S=2\pi^2 \int_0^1 \mathrm{d} t \left[N\left(4-\frac{q}{p} - p\Lambda \right) +\frac{1}{4N}\left(-\frac{q}{p}\dot{p}^2 + 4 q \ddot{p} + 2 p \ddot{q} + 4 \dot{p} \dot{q}\right)\right] \,.
\end{align}
Variation of the action leads to the equations of motion
\begin{align}
    2p\ddot{p} -\dot{p}^2 & = 4N^2\,, \label{eomp} \\
    \ddot{q} + \frac{\dot{p}}{p}\dot{q} & = N^2 \left(2\Lambda - 4\frac{q}{p^2} \right)\,,
\end{align}
and the (Neumann) boundary variations
\begin{align}
    \left( q \delta \Pi_q + p \delta \Pi_p \right)\mid_{t=1} - \left( q \delta \Pi_q + p \delta \Pi_p\right)\mid_{t=0}\,, \label{bdyvariation}
\end{align}
which are expressed in terms of the canonical momenta
\begin{align}
    \frac{1}{2\pi^2}\Pi_q = -\frac{1}{2N}\dot{p}\,, \qquad \frac{1}{2\pi^2}\Pi_p = -\frac{1}{2N}\left( \dot{q} + \frac{q}{p}\dot{p}\right)\,.
\end{align}
Note that the action is linear in $q.$ This means that when the path integral over $q$ is performed, it will give rise to a functional delta function, imposing the equation of motion \eqref{eomp} and eliminating $q$. We choose the boundary conditions again to impose regularity. However, we must be careful not to impose conditions on a variable and its conjugate momentum simultaneously (which would violate the commutation relations). An appropriate choice is
\begin{align}
    & \Pi_q|_{t=0} = -2\pi^2i\,, \qquad p|_{t=0}=0\,, \\
    & \Pi_q|_{t=1} = +2\pi^2i\,, \qquad p|_{t=1}=0\,.
\end{align}
With these boundary conditions, the equations of motion are solved by
\begin{align}
    p(t) = 2Ni(t-t^2)\,, \qquad q(t) = -\frac{\Lambda}{3}N^2 (t-t^2)\,, 
\end{align}
and the action reduces to a function of the lapse only,
\begin{align}
    S_0(N) = 2\pi^2\left( -\frac{\Lambda}{3}iN^2 + 4N\right)\,.
\end{align}
There is just a single saddle point, located at
\begin{align}
    N_\sigma = - \frac{6}{\Lambda}i\,, \qquad iS_0(N_\sigma) = +\frac{24\pi^2}{\Lambda}\,.
\end{align}
Note that at the saddle point, $p(t)=q(t)=\frac{12}{\Lambda}(t-t^2),$ and the saddle point corresponds to the round $4$-sphere.

\begin{figure}[htbp]
\centering
\includegraphics[width=0.4\textwidth]{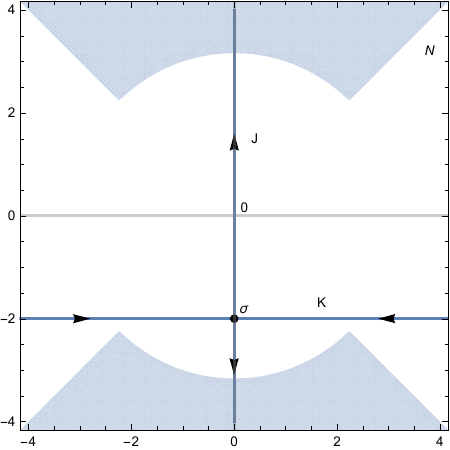}
\caption{Saddle point $\sigma$ and associated flow lines in the complex plane of the lapse function, for Neumann-Neumann boundary conditions imposing regularity of the metric at both ends. The shaded areas indicate the asymptotic regions of convergence of the path integral. Here $\Lambda=3.$}  \label{fig:NN}
\end{figure}

Fig.~\ref{fig:NN} shows the flow lines associated with the saddle point $N_\sigma$ in the complex $N$ plane. There is a unique Lefschetz thimble, and it runs along the Euclidean lapse axis. The two regions of convergence are the wedges between angles $\pi/4$ and $3\pi/4,$ and also between $5\pi/4$ and $7\pi/4.$ Thus there is a single equivalence class of convergent contours, and they are all equivalent to the Euclidean contour. Hence, in this case, we must choose the Euclidean contour of integration -- that is the only possibility for defining the path integral. The vacuum persistence (with Neumann boundary conditions) thus stands out as a special case in which the integration over Euclidean lapse leads to a convergent integral. It is dominated by the saddle point that is a sphere, and that has the expected enhanced weighting,
\begin{align}
    \langle \varnothing_N | \varnothing_N \rangle \approx e^{24\pi^2/\Lambda}\,.
\end{align}
As discussed previously, such sphere contributions could be added to all no-boundary amplitudes by splitting the time segment into arbitrarily many pieces, and this would yield pathological necklace configurations such as the one shown in Fig.~\ref{fig:necklace}. Here also we see that the specification of boundary conditions and integration contours is not enough, and the additional restriction to connected and allowable geometries is required in order to obtain physically sensible results.

\section{Discussion}

Feynman's sum-over-histories approach to quantum mechanics not only provided a mathematically useful alternative to the older Heisenberg and Schr\"{o}dinger methods (and proved especially useful for incorporating non-perturbative effects in its generalization to quantum field theory), but it also provided a conceptually new set of ideas that allowed a straightforward incorporation of gravity, simply by adding the Einstein-Hilbert term to the total covariantized action. But this generalization is not quite as straightforward as one might think at first. In quantum mechanics, the path integral is performed over all paths satisfying the boundary conditions. The dependence on boundary conditions remains crucial when gravity is included, but that is by far not enough. The generalization from ``all paths'' to ``all geometries'' is known not to work, as many pathological geometries exist that render the path integral either mathematically or physically inconsistent (or both). 

Presently, we are still quite far from a full understanding of exactly which geometries should be included. Mathematical consistency eliminates many possibilities -- in particular it removes Euclidean contours of the lapse in most situations (except for the vacuum persistence, as we saw).
Physical consistency is often achieved by imposing a Lorentzian contour, which is not surprising given the successes of relativistic physics and the general importance of causality. The no-boundary proposal can then be implemented via a regularity condition on the momentum. The fact that this condition requires the momentum to be imaginary fits well with the intuition that purely real Lorentzian solutions describe classical physics, while complex solutions can capture intrinsically quantum effects. 

Boundary conditions and integration contours have been much discussed in the past, but as we reviewed, their specification is not sufficient for obtaining physically consistent results. The main obstruction to obtaining sensible results comes from contributions of disconnected geometries. Let us stress that it is important to find a proper normalization of the gravitational wave functions: if it does not hold, then the Schwarz inequality will not hold either, and as a consequence the Heisenberg uncertainty relation will not be satisfied. Hence the normalization is crucial for the physical consistency of the whole approach. The obstruction coning from disconnected geometries then suggests a remedy: restrict to connected geometries only.

This restriction gains support from the KSW allowability criterion for complex geometries. Indeed, regularized disconnected geometries always strongly violate this criterion (hence it is quite conceivable that a somewhat weaker condition would do -- this is an interesting question for future work). But the main motivation for connected geometries comes from a focus on observable quantities: spacetimes that are not connected to ours appear to be unobservable in principle, and should thus not be part of our physical theories.

Such a restriction has important consequences: first of all, it immediately eliminates (even from very mundane transitions) bizarre contributions in which the universe shrinks to nothing and re-emerges. Moreover, in no-boundary amplitudes, the vacuum persistence does not factor out anymore. It also does not exponentiate, but in fact serves to normalize the amplitudes by providing a common factor that receives its main contribution from the lowest vacuum in the theory (in this context see also the recent discussion of the lowest dS vacuum as a state of maximal entropy in \cite{Witten:2026whn}). 

Restricting to connected geometries runs somewhat contrary to recent discussions of factorization in AdS/CFT, especially in situations involving two disconnected boundary CFTs \cite{Maldacena:2001kr}. It is important to keep in mind that in that setup, the two boundary CFTs are spacelike separated, and thus a fair amount of ``disconnectedness'' is built into it. In particular, one does not consider causal processes by which the boundaries would be created and then separated, via physical operations, to their eventual locations. Rather, one might consider this setup as a toy model that is useful to study certain mathematical aspects of quantum gravity, but less so a realistic model of our universe.

There are many avenues for future work, in addition to those mentioned previously. The most important one is to include a scalar field, allowing for a realistic inflationary phase, and a phase of dark energy domination. More generally, one may wonder how different matter types affect the normalization of transition amplitudes. In addition, it would be interesting to include wormholes in the analysis (these are certainly allowed, as they are purely Euclidean and connect distant spacetime regions). The most relevant examples might be the wormholes of wineglass shape, which can nucleate a baby universe undergoing an inflationary phase \cite{Lavrelashvili:1987jg,Lavrelashvili:1988un,Hebecker:2018ofv,Betzios:2026rbv}. Moreover, such wormholes have recently been found to be closely related to no-boundary instantons, but with a weighting that is always suppressed compared to the latter \cite{Lavrelashvili:2026zsw,Lavrelashvili:2026jcw,Lavrelashvili:2026nib}. These might indeed have a nontrivial effect on the normalization of gravitational wave functions.

\begin{center}
    {\bf Acknowledgements}
\end{center}
 
I would like to thank Luca Iliesiu for correspondence and George Lavrelashvili for comments on the manuscript. I thank the Albert-Einstein-Institute for kind hospitality while this work was conducted.

\appendix

\section{Normalization in quantum mechanics} \label{sec:norm}

Quantum mechanics deals with states that are modeled as rays in a Hilbert space. A ray is an ensemble of normalized vectors that differ only by multiplication with a complex number. This prompts us to briefly review how a normalized vector is constructed. Start with any vector $|\psi\rangle$ (where we adopt Dirac's bracket notation). We can express it as a linear combination of (linearly independent, orthogonal) basis vectors $|\phi_i\rangle,$ where in a $d$-dimensional Hilbert space $i=1,\dots,d$ and $d$ can be infinite,
\begin{align}
    |\psi\rangle = \sum_{i=1}^d \alpha_i |\phi_i\rangle \qquad \rightarrow \qquad \alpha_j = \frac{\langle \phi_j | \psi \rangle}{\langle \phi_j|\phi_j \rangle}\,,
\end{align}
where the $\alpha_i$ are complex numbers and we made use of the inner product $\langle|\rangle$ that the Hilbert space is equipped with (as well as orthogonality of the $|\phi_i\rangle$). Putting the two relations together, we obtain an expression of a vector in terms of the basis vectors,
\begin{align}
    |\psi\rangle = \sum_{i=1}^d \frac{\langle \phi_i | \psi \rangle}{\langle \phi_i|\phi_i \rangle} |\phi_i\rangle\,.
\end{align}
We can use this expression to write the scalar product of two vectors in terms of the basis vectors, as
\begin{align}
    \langle \psi_1 | \psi_2 \rangle = \sum_{i,j}\frac{\langle \psi_1|\phi_j\rangle}{\langle \phi_j|\phi_j \rangle}\frac{\langle \phi_i|\psi_2\rangle}{\langle\phi_i |\phi_i \rangle}\langle \phi_j|\phi_i \rangle = \sum_i \frac{\langle \psi_1|\phi_i \rangle \langle \phi_i |\psi_2 \rangle}{\langle \phi_i|\phi_i\rangle}\,.
\end{align}
From this, one defines the probability of reaching the state $|\phi_i\rangle$ from $|\psi\rangle$ as
\begin{align}
    P_{|\psi\rangle \to |\phi_i\rangle} \equiv \frac{|\langle \phi_i| \psi \rangle|^2}{\langle \phi_i|\phi_i \rangle \langle \psi | \psi \rangle}\,.\label{qmprob}
\end{align}
Note that these probabilities are positive and sum to unity as a consequence of the properties of the inner product. We could also work with normalized states, by rescaling $|\Psi\rangle = \frac{|\psi\rangle}{\sqrt{\langle \psi | \psi \rangle}}.$

The so-called Schwarz inequality guarantees that the numerator in \eqref{qmprob} is always smaller than or equal to the denominator. From this property follows the Heisenberg uncertainty relation.

\section{Closed de Sitter space} \label{sec:dS}

The most symmetric solution to a theory of gravity with positive cosmological constant $\Lambda>0,$ with action 
\begin{align}
    S = \frac{1}{2}\int \mathrm{d}^4 x \sqrt{-g} \left( R - 2\Lambda \right)\,,
\end{align}
is de Sitter spacetime. The spatial sections can be taken to be open, flat or closed. Of particular interest in the present context is the closed slicing, in which the spatial sections are $3$-spheres with metric $\mathrm{d}\Omega_3^2.$ The metric is then given by
\begin{align}
    \mathrm{d}s^2 = - \mathrm{d}t^2 + \frac{1}{H^2}\cosh^2(Ht) \, \mathrm{d}\Omega_3^2\,,
\end{align}
where $t$ is the time coordinate and $H=\sqrt{\Lambda/3}.$ The scale factor is consequently $a(t)=\cosh(Ht)/H.$ The de Sitter (dS) solutions can be pictured as a hyperboloid, when embedded into a higher-dimensional space. Note that the scale factor has a minimum value of $a(t=0)=1/H$ at the waist. For values larger than $1/H,$ there are always two possible values of the time coordinate, since $\cosh(-x)=\cosh(x).$

We are also interested in the Euclidean time version of this solution. Analytically continuing $t = -i(\tau - \pi/2H),$ we obtain 
\begin{align}
    \mathrm{d}s^2 = + \mathrm{d}\tau^2 + \frac{1}{H^2}\sin^2(H\tau) \, \mathrm{d}\Omega_3^2\,,
\end{align}
This is the metric on a $4$-sphere. Now the scale factor $a(\tau) = \sin(H\tau)/H$ takes all values between zero and a maximum value $a(\tau=\pi/2H)=1/H$ at the equator. Again, each of these values is reached twice, once on each side of the equator.

The scale factor value $a=1/H$ is the maximum of the Euclidean solution and the minimum of the Lorentzian one. Moreover, the time derivative of the scale factor vanishes there. One can thus glue half of the $4$-sphere to a portion of the Lorentzian hyperboloid, by joining the two together at the equator/waist. This results in the prototypical no-boundary instanton.

The action along the Euclidean hemisphere is purely imaginary, and gives a non-trivial weighting to the no-boundary solution. Along the Lorentzian part of the solution, the action is real and contributes only a phase. If the instanton reaches final scale factor $a_f,$ the action is given by
\begin{align}
    S_{\mathrm{no-boundary}} = -i \frac{12\pi^2}{\Lambda} - \frac{12\pi^2}{\Lambda}\left(\frac{\Lambda}{3} a_f^2 -1\right)^{3/2}
\end{align}
For a full sphere, which appears in the vacuum persistence, one obtains simply twice the result for the hemisphere, namely 
\begin{align}
    S_{4\mathrm{-sphere}} = -i \frac{24\pi^2}{\Lambda}\,.
\end{align}

\section{A central property of Picard-Lefshetz theory} \label{sec:Picard}

The Feynman path integral in quantum mechanics as well as its generalizations used in quantum gravity involve oscillating integrals of the form 
\begin{align}
\label{eq:integral}
I = \int_{\cal C} \mathrm{d}x\, e^{\frac{i}{\hbar}S(x)}\,,
\end{align}
where $S(x)$ is a real function (the action), $\hbar$ a real parameter, and ${\cal C}$ an integration contour. Such integrals can be difficult to evaluate, precisely because of the oscillations. Picard-Lefschetz (PL) theory provides a way of defining such integrals unambiguously, and of evaluating (or at least approximating) them much more easily. It achieves this by analytically continuing the integral to the complex plane, and deforming the integration contour in an optimal way, so as to remove the oscillations. Integration contours can also be intrinsically complex, and PL theory allows one to deal with these equally straightforwardly. For a much more complete treatment, see \cite{Witten:2010cx,Feldbrugge:2017kzv}. Here we will focus on a single question, namely how to determine which saddle points contribute to a given integral.

We first extend the variable $x$ to the complex plane, $x  \in \mathbb{C},$ and view $S(x)$ as a holomorphic function of $x.$ It is useful to write $x=u^1+i u^2$ for real $u^1, u^2.$ The integrand is decomposed as  $\mathcal{I}=iS/\hbar = W + i P,$ where $W$ is the \emph{weighting} (sometimes called Morse function) and $P$ the \emph{phase}. Downward flow (by which we mean a flow in the direction of decreasing weighting) along a curve with parameter $\lambda$ is defined by
\begin{equation}
\frac{\mathrm{d}u^i}{\mathrm{d}\lambda} = -g^{ij}\frac{\partial W}{\partial u^j}\,,
\label{eq:dw}
\end{equation}
where $g_{ij}$ is a (Riemannian) metric on the complex plane. Along the flow
\begin{align}
\frac{\mathrm{d}W}{\mathrm{d} \lambda} = \sum_i\frac{\partial W}{\partial u^i}\frac{\mathrm{d}u^i}{\mathrm{d}\lambda} = -\sum_i\left(\frac{\partial W}{\partial u^i}\right)^2 \leq 0\,,
\end{align}
confirming that it is decreasing (in fact it is decreasing as fast as possible along such lines -- one can show that the phase $P$ is constant along them). If there is a stationary point of the weighting, the Cauchy-Riemann equations imply that it must be a saddle point ($\partial S/\partial x = 0$) of the full integrand. The downwards flows emanating from saddle points are called  {\it Lefschetz thimbles}, and the integral converges most rapidly along such thimbles. In the same spirit one may define upwards flows, via
\begin{equation}
\frac{\mathrm{d}u^i}{\mathrm{d}\lambda} = +g^{ij}\frac{\partial W}{\partial u^j}\,.
\label{eq:up}
\end{equation}

We are now ready to address the question of the relevance of saddle points. Let us assume that to each saddle point $\sigma$ we have a single associated steepest descent contour ${\cal J}_\sigma$, and a single steepest ascent contour ${\cal K}_\sigma$ (if there are degeneracies, one can eliminate them via a small complex addition to the action, which is then removed at the end of the calculation). Mathematically, we can write this as an intersection
\begin{equation}
{\rm Int}({\cal J}_\sigma, {\cal K}_{\sigma'})=\delta_{\sigma \sigma'}\,, \label{eq:intersection}
\end{equation}
where a direction is implicitly assigned to each thimble and steepest ascent line. A central idea is to rewrite the integration contour ${\cal C}$ of the integral as a sum over thimbles, 
\begin{equation}
\label{eq:contourexp}
{\cal C} = \sum_\sigma n_\sigma {\cal J}_\sigma\,, 
\end{equation}
with $n_\sigma \in \{ 0, \pm 1\}.$ For the integrals of interest, such a rewriting can always be done. Then we can extract the $n_\sigma$ coefficients by multiplying both sides with an ascent contour and taking the intersection,
\begin{align}
n_\sigma= {\rm Int}(\mathcal{C}, {\cal K}_{\sigma})\,. \label{tocontributeornot}
\end{align}
This provides the desired criterion: a particular thimble (and its associated saddle point) contributes to the integral if the original integration contour intersects its steepest ascent contour. This makes perfect sense, as there are no oscillations of the integrand along a thimble, and hence the weighting along a thimble (and in particular at a saddle point) must be lower than that along the original integration contour. There is an important special case: if the action is real and the original integration contour Lorentzian, so that the weighting along the original contour is zero, then only saddle points with a negative/suppressed weighting may contribute to the integral.

In a very direct manner, PL theory allows one to perform a saddle point approximation of the integral,
\begin{align}
I = \int_{\cal C} \mathrm{d}x \, e^{i S[x]/\hbar}  = \sum_\sigma n_\sigma \int_{{\cal J}_\sigma} \mathrm{d} x \, e^{\frac{i}{\hbar}S[x]} 
 \approx \sum_\sigma n_\sigma \, e^{i S(x_\sigma)/\hbar}\,.
\end{align} 
Here we wrote the integral as a sum of integrals over thimbles, and then used the fact that along a thimble the largest contribution occurs at the saddle point $\sigma$ (assuming $\hbar$ is small).

\renewcommand{\baselinestretch}{1.21}
\bibliographystyle{utphys}
\bibliography{biblio}

@book{Misner:1973prb,
    author = "Misner, Charles W. and Thorne, K. S. and Wheeler, J. A.",
    title = "{Gravitation}",
    isbn = "978-0-7167-0344-0, 978-0-691-17779-3",
    publisher = "W. H. Freeman",
    address = "San Francisco",
    year = "1973"
}

@article{Hartle:1983ai,
    author = "Hartle, J. B. and Hawking, S. W.",
    editor = "Fang, Li-Zhi and Ruffini, R.",
    title = "{Wave Function of the Universe}",
    reportNumber = "PRINT-83-0937 (CAMBRIDGE)",
    doi = "10.1103/PhysRevD.28.2960",
    journal = "Phys. Rev. D",
    volume = "28",
    pages = "2960--2975",
    year = "1983"
}

@article{Lavrelashvili:1987jg,
    author = "Lavrelashvili, G. and Rubakov, V. A. and Tinyakov, P. G.",
    title = "{Disruption of Quantum Coherence upon a Change in Spatial Topology in Quantum Gravity}",
    journal = "JETP Lett.",
    volume = "46",
    pages = "167--169",
    year = "1987"
}

@article{Lavrelashvili:1988un,
    author = "Lavrelashvili, G. and Rubakov, V. A. and Tinyakov, P. G.",
    title = "{Loss of Quantum Coherence Due to Topological Changes: A Toy Model}",
    doi = "10.1142/S0217732388001483",
    journal = "Mod. Phys. Lett. A",
    volume = "3",
    pages = "1231--1242",
    year = "1988"
}

@article{Hebecker:2018ofv,
    author = "Hebecker, Arthur and Mikhail, Thomas and Soler, Pablo",
    title = "{Euclidean wormholes, baby universes, and their impact on particle physics and cosmology}",
    eprint = "1807.00824",
    archivePrefix = "arXiv",
    primaryClass = "hep-th",
    doi = "10.3389/fspas.2018.00035",
    journal = "Front. Astron. Space Sci.",
    volume = "5",
    pages = "35",
    year = "2018"
}

@article{Loges:2022nuw,
    author = "Loges, Gregory J. and Shiu, Gary and Sudhir, Nidhi",
    title = "{Complex saddles and Euclidean wormholes in the Lorentzian path integral}",
    eprint = "2203.01956",
    archivePrefix = "arXiv",
    primaryClass = "hep-th",
    doi = "10.1007/JHEP08(2022)064",
    journal = "JHEP",
    volume = "08",
    pages = "064",
    year = "2022"
}

@article{Feldbrugge:2017mbc,
    author = "Feldbrugge, Job and Lehners, Jean-Luc and Turok, Neil",
    title = "{No rescue for the no boundary proposal: Pointers to the future of quantum cosmology}",
    eprint = "1708.05104",
    archivePrefix = "arXiv",
    primaryClass = "hep-th",
    doi = "10.1103/PhysRevD.97.023509",
    journal = "Phys. Rev. D",
    volume = "97",
    number = "2",
    pages = "023509",
    year = "2018"
}

@article{Feldbrugge:2017kzv,
    author = "Feldbrugge, Job and Lehners, Jean-Luc and Turok, Neil",
    title = "{Lorentzian Quantum Cosmology}",
    eprint = "1703.02076",
    archivePrefix = "arXiv",
    primaryClass = "hep-th",
    doi = "10.1103/PhysRevD.95.103508",
    journal = "Phys. Rev. D",
    volume = "95",
    number = "10",
    pages = "103508",
    year = "2017"
}

@article{Feldbrugge:2018gin,
    author = "Feldbrugge, Job and Lehners, Jean-Luc and Turok, Neil",
    title = "{Inconsistencies of the New No-Boundary Proposal}",
    eprint = "1805.01609",
    archivePrefix = "arXiv",
    primaryClass = "hep-th",
    doi = "10.3390/universe4100100",
    journal = "Universe",
    volume = "4",
    number = "10",
    pages = "100",
    year = "2018"
}

@article{Lehners:2023yrj,
    author = "Lehners, Jean-Luc",
    title = "{Review of the no-boundary wave function}",
    eprint = "2303.08802",
    archivePrefix = "arXiv",
    primaryClass = "hep-th",
    doi = "10.1016/j.physrep.2023.06.002",
    journal = "Phys. Rept.",
    volume = "1022",
    pages = "1--82",
    year = "2023"
}

@article{Betzios:2026rbv,
    author = "Betzios, Panos and Ghiringhelli, Paul and Gialamas, Ioannis D. and Papadoulaki, Olga",
    title = "{A menagerie of wormholes and cosmologies in the gravitational path integral}",
    eprint = "2602.23432",
    archivePrefix = "arXiv",
    primaryClass = "hep-th",
    doi = "10.1007/JHEP06(2026)171",
    journal = "JHEP",
    volume = "06",
    pages = "171",
    year = "2026"
}

@article{Lyons:1992ua,
    author = "Lyons, G. W.",
    title = "{Complex solutions for the scalar field model of the universe}",
    doi = "10.1103/PhysRevD.46.1546",
    journal = "Phys. Rev. D",
    volume = "46",
    pages = "1546--1550",
    year = "1992"
}

@article{Witten:2021nzp,
    author = "Witten, Edward",
    title = "{A Note On Complex Spacetime Metrics}",
    eprint = "2111.06514",
    archivePrefix = "arXiv",
    primaryClass = "hep-th",
    month = "11",
    year = "2021"
}

@article{Hertog:2023vot,
    author = "Hertog, Thomas and Janssen, Oliver and Karlsson, Joel",
    title = "{Kontsevich-Segal Criterion in the No-Boundary State Constrains Inflation}",
    eprint = "2305.15440",
    archivePrefix = "arXiv",
    primaryClass = "hep-th",
    doi = "10.1103/PhysRevLett.131.191501",
    journal = "Phys. Rev. Lett.",
    volume = "131",
    number = "19",
    pages = "191501",
    year = "2023"
}

@article{Lehners:2021mah,
    author = "Lehners, Jean-Luc",
    title = "{Allowable complex metrics in minisuperspace quantum cosmology}",
    eprint = "2111.07816",
    archivePrefix = "arXiv",
    primaryClass = "hep-th",
    doi = "10.1103/PhysRevD.105.026022",
    journal = "Phys. Rev. D",
    volume = "105",
    number = "2",
    pages = "026022",
    year = "2022"
}

@article{Jonas:2022uqb,
    author = "Jonas, Caroline and Lehners, Jean-Luc and Quintin, Jerome",
    title = "{Uses of complex metrics in cosmology}",
    eprint = "2205.15332",
    archivePrefix = "arXiv",
    primaryClass = "hep-th",
    doi = "10.1007/JHEP08(2022)284",
    journal = "JHEP",
    volume = "08",
    pages = "284",
    year = "2022"
}

@article{Lehners:2022xds,
    author = "Lehners, Jean-Luc",
    title = "{Allowable complex scalars from Kaluza-Klein compactifications and metric rescalings}",
    eprint = "2209.14669",
    archivePrefix = "arXiv",
    primaryClass = "hep-th",
    doi = "10.1103/PhysRevD.107.046004",
    journal = "Phys. Rev. D",
    volume = "107",
    number = "4",
    pages = "046004",
    year = "2023"
}

@article{Lehners:2023pcn,
    author = "Lehners, Jean-Luc and Quintin, Jerome",
    title = "{A small Universe}",
    eprint = "2309.03272",
    archivePrefix = "arXiv",
    primaryClass = "hep-th",
    doi = "10.1016/j.physletb.2024.138488",
    journal = "Phys. Lett. B",
    volume = "850",
    pages = "138488",
    year = "2024"
}

@article{Lavrelashvili:2026zsw,
    author = "Lavrelashvili, George and Lehners, Jean-Luc",
    title = "{Nucleating an Inflationary Universe: Euclidean Wormholes and their No-Boundary Limit}",
    eprint = "2603.11003",
    archivePrefix = "arXiv",
    primaryClass = "hep-th",
    month = "3",
    year = "2026"
}

@article{Abdalla:2026mxn,
    author = "Abdalla, Ahmed I. and Antonini, Stefano and Bousso, Raphael and Iliesiu, Luca V. and Levine, Adam and Shahbazi-Moghaddam, Arvin",
    title = "{Consistent Evaluation of the No-Boundary Proposal}",
    eprint = "2602.02682",
    archivePrefix = "arXiv",
    primaryClass = "hep-th",
    month = "2",
    year = "2026"
}

@article{Kontsevich:2021dmb,
    author = "Kontsevich, Maxim and Segal, Graeme",
    title = "{Wick Rotation and the Positivity of Energy in Quantum Field Theory}",
    eprint = "2105.10161",
    archivePrefix = "arXiv",
    primaryClass = "hep-th",
    doi = "10.1093/qmath/haab027",
    journal = "Quart. J. Math. Oxford Ser.",
    volume = "72",
    number = "1-2",
    pages = "673--699",
    year = "2021"
}

@article{Lavrelashvili:2026jcw,
    author = "Lavrelashvili, George and Lehners, Jean-Luc",
    title = "{Birth of inflationary universes via wineglass wormholes and their no-boundary relatives}",
    eprint = "2605.10548",
    archivePrefix = "arXiv",
    primaryClass = "hep-th",
    doi = "10.1103/jpwp-2nzv",
    journal = "Phys. Rev. D",
    volume = "114",
    number = "4",
    pages = "046009",
    year = "2026"
}

@article{Lavrelashvili:2026nib,
    author = "Lavrelashvili, George and Lehners, Jean-Luc",
    title = "{Quantum States Prepared by Wormholes: Long-Wavelength Deviations from Bunch-Davies}",
    eprint = "2607.12772",
    archivePrefix = "arXiv",
    primaryClass = "hep-th",
    month = "7",
    year = "2026"
}

@article{Honda:2024aro,
    author = "Honda, Masazumi and Matsui, Hiroki and Okabayashi, Kazumasa and Terada, Takahiro",
    title = "{Resurgence in Lorentzian quantum cosmology: no-boundary saddles and resummation of quantum gravity corrections around tunneling saddles}",
    eprint = "2402.09981",
    archivePrefix = "arXiv",
    primaryClass = "gr-qc",
    reportNumber = "RIKEN-iTHEMS-Report-24, YITP-24-10, CTPU-PTC-24-04",
    month = "2",
    year = "2024"
}

@article{Ailiga:2023wzl,
    author = "Ailiga, Manishankar and Mallik, Shubhashis and Narain, Gaurav",
    title = "{Lorentzian Robin Universe}",
    eprint = "2308.01310",
    archivePrefix = "arXiv",
    primaryClass = "gr-qc",
    doi = "10.1007/JHEP01(2024)124",
    journal = "JHEP",
    volume = "01",
    pages = "124",
    year = "2024"
}

@article{Matsui:2022lfj,
    author = "Matsui, Hiroki and Mukohyama, Shinji and Naruko, Atsushi",
    title = "{No smooth spacetime in Lorentzian quantum cosmology and trans-Planckian physics}",
    eprint = "2211.05306",
    archivePrefix = "arXiv",
    primaryClass = "gr-qc",
    reportNumber = "YITP-22-133, IPMU22-0058",
    doi = "10.1103/PhysRevD.107.043511",
    journal = "Phys. Rev. D",
    volume = "107",
    number = "4",
    pages = "043511",
    year = "2023"
}

@inbook{Hawking:1980gf,
    author = "Hawking, S. W.",
    title = "{The Path Integral Approach To Quantum Gravity}",
    booktitle = "{General Relativity}: {An Einstein Centenary Survey}",
    pages = "746--789",
    year = "1980"
}

@article{Halliwell:1988ik,
    author = "Halliwell, Jonathan J. and Louko, Jorma",
    title = "{Steepest Descent Contours in the Path Integral Approach to Quantum Cosmology. 1. The De Sitter Minisuperspace Model}",
    reportNumber = "NSF-ITP-88-173",
    doi = "10.1103/PhysRevD.39.2206",
    journal = "Phys. Rev. D",
    volume = "39",
    pages = "2206",
    year = "1989"
}

@article{Krishnan:2016mcj,
    author = "Krishnan, Chethan and Raju, Avinash",
    title = "{A Neumann Boundary Term for Gravity}",
    eprint = "1605.01603",
    archivePrefix = "arXiv",
    primaryClass = "hep-th",
    doi = "10.1142/S0217732317500778",
    journal = "Mod. Phys. Lett. A",
    volume = "32",
    number = "14",
    pages = "1750077",
    year = "2017"
}

@article{DiTucci:2020weq,
    author = "Di Tucci, Alice and Heller, Michal P. and Lehners, Jean-Luc",
    title = "{Lessons for quantum cosmology from anti\textendash{}de Sitter black holes}",
    eprint = "2007.04872",
    archivePrefix = "arXiv",
    primaryClass = "hep-th",
    doi = "10.1103/PhysRevD.102.086011",
    journal = "Phys. Rev. D",
    volume = "102",
    number = "8",
    pages = "086011",
    year = "2020"
}

@article{Witten:2010cx,
    author = "Witten, Edward",
    editor = "Andersen, Joergen E. and Boden, Hans U. and Hahn, Atle and Himpel, Benjamin",
    title = "{Analytic Continuation Of Chern-Simons Theory}",
    eprint = "1001.2933",
    archivePrefix = "arXiv",
    primaryClass = "hep-th",
    journal = "AMS/IP Stud. Adv. Math.",
    volume = "50",
    pages = "347--446",
    year = "2011"
}

@article{Lehners:2019ibe,
    author = "Lehners, Jean-Luc and Stelle, K. S.",
    title = "{A Safe Beginning for the Universe?}",
    eprint = "1909.01169",
    archivePrefix = "arXiv",
    primaryClass = "hep-th",
    reportNumber = "Imperial/TP/2019/KSS/02",
    doi = "10.1103/PhysRevD.100.083540",
    journal = "Phys. Rev. D",
    volume = "100",
    number = "8",
    pages = "083540",
    year = "2019"
}

@article{Hertog:2013mra,
    author = "Hertog, Thomas",
    title = "{Predicting a Prior for Planck}",
    eprint = "1305.6135",
    archivePrefix = "arXiv",
    primaryClass = "astro-ph.CO",
    doi = "10.1088/1475-7516/2014/02/043",
    journal = "JCAP",
    volume = "02",
    pages = "043",
    year = "2014"
}

@article{DiazDorronsoro:2017hti,
    author = "Diaz Dorronsoro, Juan and Halliwell, Jonathan J. and Hartle, James B. and Hertog, Thomas and Janssen, Oliver",
    title = "{Real no-boundary wave function in Lorentzian quantum cosmology}",
    eprint = "1705.05340",
    archivePrefix = "arXiv",
    primaryClass = "gr-qc",
    doi = "10.1103/PhysRevD.96.043505",
    journal = "Phys. Rev. D",
    volume = "96",
    number = "4",
    pages = "043505",
    year = "2017"
}

@article{Janssen:2019sex,
    author = "Janssen, Oliver and Halliwell, Jonathan J. and Hertog, Thomas",
    title = "{No-boundary proposal in biaxial Bianchi IX minisuperspace}",
    eprint = "1904.11602",
    archivePrefix = "arXiv",
    primaryClass = "gr-qc",
    doi = "10.1103/PhysRevD.99.123531",
    journal = "Phys. Rev. D",
    volume = "99",
    number = "12",
    pages = "123531",
    year = "2019"
}

@article{DiTucci:2019bui,
    author = "Di Tucci, Alice and Lehners, Jean-Luc and Sberna, Laura",
    title = "{No-boundary prescriptions in Lorentzian quantum cosmology}",
    eprint = "1911.06701",
    archivePrefix = "arXiv",
    primaryClass = "hep-th",
    doi = "10.1103/PhysRevD.100.123543",
    journal = "Phys. Rev. D",
    volume = "100",
    number = "12",
    pages = "123543",
    year = "2019"
}

@article{DiTucci:2019dji,
    author = "Di Tucci, Alice and Lehners, Jean-Luc",
    title = "{No-Boundary Proposal as a Path Integral with Robin Boundary Conditions}",
    eprint = "1903.06757",
    archivePrefix = "arXiv",
    primaryClass = "hep-th",
    doi = "10.1103/PhysRevLett.122.201302",
    journal = "Phys. Rev. Lett.",
    volume = "122",
    number = "20",
    pages = "201302",
    year = "2019"
}

@article{Halliwell:1989vu,
    author = "Halliwell, Jonathan J. and Louko, Jorma",
    title = "{Steepest Descent Contours in the Path Integral Approach to Quantum Cosmology. 2. Microsuperspace}",
    reportNumber = "NSF-ITP-89-21",
    doi = "10.1103/PhysRevD.40.1868",
    journal = "Phys. Rev. D",
    volume = "40",
    pages = "1868",
    year = "1989"
}

@article{Halliwell:1990tu,
    author = "Halliwell, Jonathan J. and Louko, Jorma",
    title = "{Steepest Descent Contours in the Path Integral Approach to Quantum Cosmology. 3. A General Method With Applications to Anisotropic Minisuperspace Models}",
    reportNumber = "MIT-CTP-1846",
    doi = "10.1103/PhysRevD.42.3997",
    journal = "Phys. Rev. D",
    volume = "42",
    pages = "3997--4031",
    year = "1990"
}

@article{DiazDorronsoro:2018wro,
    author = "Diaz Dorronsoro, Juan and Halliwell, Jonathan J. and Hartle, James B. and Hertog, Thomas and Janssen, Oliver and Vreys, Yannick",
    title = "{Damped perturbations in the no-boundary state}",
    eprint = "1804.01102",
    archivePrefix = "arXiv",
    primaryClass = "gr-qc",
    doi = "10.1103/PhysRevLett.121.081302",
    journal = "Phys. Rev. Lett.",
    volume = "121",
    number = "8",
    pages = "081302",
    year = "2018"
}

@article{Narain:2022msz,
    author = "Narain, Gaurav",
    title = "{Surprises in Lorentzian path-integral of Gauss-Bonnet gravity}",
    eprint = "2203.05475",
    archivePrefix = "arXiv",
    primaryClass = "gr-qc",
    doi = "10.1007/JHEP04(2022)153",
    journal = "JHEP",
    volume = "04",
    pages = "153",
    year = "2022"
}

@article{Lehners:2024kus,
    author = "Lehners, Jean-Luc",
    title = "{NUT-Bolt saddle points and Stokes phenomena in the no-boundary wave function}",
    eprint = "2402.10501",
    archivePrefix = "arXiv",
    primaryClass = "gr-qc",
    doi = "10.1103/PhysRevD.111.063503",
    journal = "Phys. Rev. D",
    volume = "111",
    number = "6",
    pages = "063503",
    year = "2025"
}

@article{Chen:2023hra,
    author = "Chen, Yiming and Ivo, Victor and Maldacena, Juan",
    title = "{Comments on the double cone wormhole}",
    eprint = "2310.11617",
    archivePrefix = "arXiv",
    primaryClass = "hep-th",
    doi = "10.1007/JHEP04(2024)124",
    journal = "JHEP",
    volume = "04",
    pages = "124",
    year = "2024"
}

@article{Honda:2026wsl,
    author = "Honda, Masazumi and Matsui, Hiroki and Numajiri, Kota and Okabayashi, Kazumasa and Terada, Takahiro",
    title = "{Dimensional (In)dependence of Lorentzian Quantum Cosmology}",
    eprint = "2609.33452",
    archivePrefix = "arXiv",
    primaryClass = "gr-qc",
    reportNumber = "RIKEN-iTHEMS-Report-26, STUPP-26-304, YITP-26-126, OCU-PHYS-631, AP-GR-215",
    month = "9",
    year = "2026"
}

@article{Assant:2026dca,
    author = "Assant, Gabriel and Litim, Daniel F. and Reichert, Manuel",
    title = "{Spectral Functions of Lorentzian Quantum Gravity}",
    eprint = "2606.19321",
    archivePrefix = "arXiv",
    primaryClass = "hep-th",
    month = "6",
    year = "2026"
}

@article{Kolanowski:2026gii,
    author = "Kolanowski, Maciej and Marolf, Donald",
    title = "{How to tame your (black hole) saddles: Lessons from the Lorentzian Gravitational Path Integral}",
    eprint = "2603.24681",
    archivePrefix = "arXiv",
    primaryClass = "hep-th",
    month = "3",
    year = "2026"
}

@article{Held:2026bbo,
    author = "Held, Jesse and Kaplan, Molly and Marolf, Donald and Wang, Zhencheng",
    title = "{Lorentzian path integrals and Jackiw-Teitelboim wormholes with imaginary scalars}",
    eprint = "2601.09932",
    archivePrefix = "arXiv",
    primaryClass = "hep-th",
    doi = "10.1007/JHEP07(2026)215",
    journal = "JHEP",
    volume = "07",
    pages = "215",
    year = "2026"
}

@article{Nomura:2025whc,
    author = "Nomura, Yasunori and Ugajin, Tomonori",
    title = "{Nonperturbative quantum gravity in a closed Lorentzian universe}",
    eprint = "2505.20390",
    archivePrefix = "arXiv",
    primaryClass = "hep-th",
    reportNumber = "RIKEN-iTHEMS-Report-25",
    doi = "10.1007/JHEP10(2025)166",
    journal = "JHEP",
    volume = "10",
    pages = "166",
    year = "2025"
}

@article{Borissova:2024txs,
    author = "Borissova, Johanna and Dittrich, Bianca and Qu, Dongxue and Schiffer, Marc",
    title = "{Spikes and spines in 4D Lorentzian simplicial quantum gravity}",
    eprint = "2407.13601",
    archivePrefix = "arXiv",
    primaryClass = "gr-qc",
    doi = "10.1007/JHEP10(2024)150",
    journal = "JHEP",
    volume = "10",
    pages = "150",
    year = "2024"
}

@article{Dittrich:2024awu,
    author = {Dittrich, Bianca and Jacobson, Ted and Padua-Arg{\"u}elles, Jos{\'e}},
    title = "{de Sitter horizon entropy from a simplicial Lorentzian path integral}",
    eprint = "2403.02119",
    archivePrefix = "arXiv",
    primaryClass = "gr-qc",
    doi = "10.1103/PhysRevD.110.046006",
    journal = "Phys. Rev. D",
    volume = "110",
    number = "4",
    pages = "046006",
    year = "2024"
}

@article{Honda:2024hdr,
    author = "Honda, Masazumi and Matsui, Hiroki and Numajiri, Kota and Okabayashi, Kazumasa",
    title = "{Jackiw-Teitelboim gravity and Lorentzian quantum cosmology}",
    eprint = "2412.20398",
    archivePrefix = "arXiv",
    primaryClass = "gr-qc",
    reportNumber = "RIKEN-iTHEMS-Report-24, STUPP-24-272, YITP-24-134",
    doi = "10.1103/vrcy-y61l",
    journal = "Phys. Rev. D",
    volume = "111",
    number = "12",
    pages = "126022",
    year = "2025"
}

@article{Lehners:2023fud,
    author = "Lehners, Jean-Luc and Stelle, K. S.",
    title = "{Higher-order gravity, finite action, and a safe beginning for the universe}",
    eprint = "2312.14048",
    archivePrefix = "arXiv",
    primaryClass = "hep-th",
    doi = "10.1140/epjp/s13360-024-05125-y",
    journal = "Eur. Phys. J. Plus",
    volume = "139",
    pages = "380",
    year = "2024"
}

@article{Mondal:2023cxx,
    author = "Mondal, Vikramaditya and Chakraborty, Sumanta",
    title = "{Lorentzian quantum cosmology with torsion}",
    eprint = "2305.01690",
    archivePrefix = "arXiv",
    primaryClass = "gr-qc",
    doi = "10.1103/PhysRevD.109.043525",
    journal = "Phys. Rev. D",
    volume = "109",
    number = "4",
    pages = "043525",
    year = "2024"
}

@article{Witten:2026whn,
    author = "Witten, Edward",
    title = "{An observer in de Sitter space, and rereading Everett}",
    eprint = "2610.04804",
    archivePrefix = "arXiv",
    primaryClass = "hep-th",
    doi = "10.1142/s0217751x26470093",
    journal = "Int. J. Mod. Phys. A",
    volume = "41",
    number = "26",
    pages = "2647009",
    year = "2026"
}

@article{Maldacena:2001kr,
    author = "Maldacena, Juan Martin",
    title = "{Eternal black holes in anti-de Sitter}",
    eprint = "hep-th/0106112",
    archivePrefix = "arXiv",
    reportNumber = "NSF-ITP-01-59",
    doi = "10.1088/1126-6708/2003/04/021",
    journal = "JHEP",
    volume = "04",
    pages = "021",
    year = "2003"
}

\end{document}